\documentclass[12pt,preprint]{aastex}

\slugcomment{ApJ in press}

\shorttitle{Luminosity Functions of Galaxy Cluster MS1054$-$0321 at $z=0.83$ based on ACS Photometry}
\shortauthors{Goto et al.}

\begin{document}

\title{The Luminosity Functions of the Galaxy Cluster MS1054$-$0321 at $z=0.83$ based on ACS Photometry}

\author{Tomotsugu Goto\altaffilmark{1},
Marc Postman\altaffilmark{2}, 
Nicholas J.G. Cross\altaffilmark{1},
G.D. Illingworth\altaffilmark{3},
K. Tran\altaffilmark{4},
D. Magee\altaffilmark{3},
M. Franx\altaffilmark{5},
N. Ben\'{\i}tez\altaffilmark{1},
R.J. Bouwens\altaffilmark{3},
R. Demarco\altaffilmark{1},
H.C. Ford\altaffilmark{1},
N.L. Homeier\altaffilmark{1},
A.R. Martel\altaffilmark{1},
F. Menanteau\altaffilmark{1},
M. Clampin\altaffilmark{6},                                                            
G.F. Hartig\altaffilmark{2},
D.R. Ardila\altaffilmark{1},
F. Bartko\altaffilmark{7}, 
J.P. Blakeslee\altaffilmark{1},
L.D. Bradley\altaffilmark{1},
T.J. Broadhurst\altaffilmark{8},
R.A. Brown\altaffilmark{2},
C.J. Burrows\altaffilmark{2},
E.S. Cheng\altaffilmark{9},
P.D. Feldman\altaffilmark{1},
D.A. Golimowski\altaffilmark{1},
C. Gronwall\altaffilmark{10},
B. Holden\altaffilmark{3},
L. Infante\altaffilmark{11},
M.J. Jee\altaffilmark{1},       
J.E. Krist\altaffilmark{2},
M.P. Lesser\altaffilmark{12},
S. Mei\altaffilmark{1},
G.R. Meurer\altaffilmark{1},
G.K. Miley\altaffilmark{5},
V. Motta\altaffilmark{1,11},
R. Overzier\altaffilmark{1,5},
M. Sirianni\altaffilmark{2,13}, 
W.B. Sparks\altaffilmark{2}, 
H.D. Tran\altaffilmark{14}, 
Z.I. Tsvetanov\altaffilmark{1},   
R.L. White\altaffilmark{1,2},
W.~Zheng\altaffilmark{1},
\& A. Zirm\altaffilmark{5}}

\altaffiltext{1}{Department of Physics and Astronomy, The Johns Hopkins
  University, 3400 North Charles Street, Baltimore, MD 21218-2686, USA.}
\altaffiltext{2}{STScI, 3700 San Martin Drive, Baltimore, MD 21218.}
\altaffiltext{3}{UCO/Lick Observatory, University of California, Santa Cruz, CA 95064.}
\altaffiltext{4}{Institute for Astronomy, ETH Hnggerberg, CH-8093 Zurich, Switzerland}
\altaffiltext{5}{Leiden Observatory, Postbus 9513, 2300 RA Leiden, Netherlands.}
\altaffiltext{6}{NASA Goddard Space Flight Center, Laboratory for Astronomy and Solar Physics, Greenbelt, MD 20771.}
\altaffiltext{7}{Bartko Science \& Technology, P.O. Box 670, Mead, CO 80542-0670.}
\altaffiltext{8}{Racah Institute of Physics, The Hebrew University, Jerusalem, Israel 91904.}
\altaffiltext{9}{Conceptual Analytics, LLC, 8209 Woburn Abbey Road, Glenn Dale, MD 20769}
\altaffiltext{10}{Department of Astronomy and Astrophysics, The Pennsylvania State University, 525 Davey Lab, University Park, PA 16802.}
\altaffiltext{11}{Departmento de Astronom\'{\i}a y Astrof\'{\i}sica, Pontificia Universidad Cat\'olica de Chile, Casilla 306, Santiago 22, Chile.}
\altaffiltext{12}{Steward Observatory, University of Arizona, Tucson, AZ 85721.}
\altaffiltext{13}{European Southern Observatory, Karl-Schwarzschild-Strasse 2, D-85748 Garching, Germany.}
\altaffiltext{14}{W. M. Keck Observatory, 65-1120 Mamalahoa Hwy., Kamuela, HI 96743}

\begin{abstract}
We present new measurements of the galaxy luminosity function (LF) and its dependence on local galaxy density, color, morphology, and clustocentric radius for the massive $z=0.83$ cluster MS1054-0321. Our analyses are based on imaging performed with the Advanced Camera for Surveys (ACS) onboard the Hubble Space Telescope (HST) in the $F606W$, $F775W$ and $F850LP$ passbands and extensive spectroscopic data obtained with the Keck Low-Resolution Imaging Spectrograph. Our main results are based on a spectroscopically selected sample of 143 cluster members with morphological classifications derived from the ACS observations.  Our three primary findings are (1) the faint-end slope of the LF is steepest in the bluest filter, (2) the LF in the inner part of the cluster (or highest density regions) has a flatter faint-end slope, and (3) the fraction of early-type galaxies is higher at the bright end of the LF, and gradually decreases toward fainter magnitudes. These characteristics are consistent with those in local galaxy clusters, indicating that, at least in massive clusters, the common characteristics of cluster LFs are established at $z=0.83$.  These results provide additional support for the hypothesis that the formation of galaxies in MS1054-0321 began at redshifts considerably greater than unity. We also find a 2$\sigma$ deficit of intrinsically faint, red galaxies ($i_{775}-z_{850}\geq 0.5$, $M_i>-19$) in this cluster. Although the significance is marginal, this trend may suggest that faint, red galaxies (which are common in $z<0.1$ rich clusters) have not yet been created in this cluster at $z=0.83$.  The giant-to-dwarf ratio in MS1054-0321 starts to increase inwards of the virial radius or when $\Sigma > 30$ Mpc$^{-2}$, coinciding with the environment where the galaxy star formation rate and the morphology-density relation start to appear.  A physical process that begins to become effective at around the virial radius or $\Sigma \sim$30 Mpc$^{-2}$ may thus be responsible for the evolution of color and luminosity of cluster galaxies.
\end{abstract}

\keywords{galaxies: clusters: general --- galaxies: clusters: individual (MS1054-0321, CL1056-03)}

\section{Introduction}\label{intro}

Butcher \& Oemler's (1978,1984) discovery that fraction of blue galaxies in clusters increases with redshift first highlighted 
the significance and strength of the evolution of the cluster galaxy population over the past 4 - 5 Gyr.
The Butcher-Oemler effect has since been confirmed by many authors (Rakos, Schombert 1995; Couch et al. 1994,1998;  Margoniner, de Carvalho 2000; Margoniner et al. 2001; Ellingson  et  al. 2001; Kodama \& Bower 2001; Goto et al. 2003a, but also see Andreon et al. 1999,2004).
High redshift clusters ($z\sim 0.9$) are also observed to have significantly larger fractions of star-forming galaxies than local clusters  (Dressler \& Gunn 1992; Postman, Lubin, \& Oke 1998; Dressler et al. 1999; Poggianti et al. 1999; Postman, Lubin, \& Oke 2001). 
Blue and star-forming galaxies are clearly more common in intermediate redshift clusters than in their local counterparts.
Furthermore, the population fraction of S0 galaxies is lower in high redshift clusters 
(Dressler et al. 1997; van Dokkum et al. 1998; Fasano et al. 2000; Jones, Smail, \& Couch 2000; Fabricant et al. 2000; also see Andreon et al. 1998;
Postman et al. 2004) and van Dokkum et al. (1998) have identified a significantly larger fraction of merging/interacting galaxies in the 
cluster MS1054-0321 at $z=0.83$.  These observations suggest that environmental processes are at least partly responsible for altering the 
morphology and star formation rates (SFR) of cluster galaxies as a function of redshift. There is likely more than one physical mechanism involved given 
the large dynamic range in density over which the morphology-density relation extends (Postman et al. 1984; Goto et al. 2003b).
However, it has been difficult to specify which specific physical processes are the dominant ones given a limited amount data
at key epochs over the past half a Hubble time.

The cluster galaxy luminosity function (LF) is a fundamental constraint on the mass assembly history of galaxy clusters
as it is related to the cluster galaxy mass function through the mass-to-light ratio. Precise and accurate measurements of 
cluster LFs as a functions of local galaxy density, clustocentric radius, galaxy structure and color 
have the potential to provide insight into the role of environmental processes 
in determining the properties of the present-day galaxy populations.  The LF is, therefore, central to many 
cosmological issues (Koo \& Kron 1992; Ostriker 1993; Binggeli et al. 1998). 

In the local universe, the cluster galaxy LFs have been studied in considerable detail 
at optical and near-infrared wavelengths (e.g., Lugger 1986; Trentham 1998; Garilli et al. 1999; Paollilo et al. 2001; 
Mart{\'{\i}}nez, Zandivarez, Merch{\' a}n, \& Dom{\'{\i}}nguez 2002; Cuesta-Bolao \& Serna 2003;De Propris et al. 2003;Lin et al. 2004). 
Goto et al. (2002b) derived the composite luminosity function of 204 clusters from the  
``SDSS Cut \& Enhance Galaxy Cluster Catalog'' (Goto et al. 2002a) in the redshift range
from $z=0.02$ to $z=0.25$ in the five SDSS bands $u^{*}$, $g^{*}$,
$r^{*}$, $i^{*}$ and $z^*$, and found  that the  faint-end slope of the luminosity function becomes flatter toward the
redder wavebands, consistent with the hypothesis that the cluster luminosity function has two distinct underlying populations; a
population of bright ellipticals with a Gaussian-like luminosity distribution that dominate the bright end, 
and a population of faint blue star-forming galaxies with a steep power-law like luminosity distribution, that dominate the faint-end.
Indeed, several studies find that cluster LFs are better described by a sum of two functions; a Gaussian for bright galaxies and a 
steep Schechter function for faint galaxies (Driver et al. 1994; Molinari, Chincarini, Moretti, \& de Grandi 1998; Parolin et al. 2003; 
Mercurio et al. 2003; Dahl{\' e}n et al. 2004), consistent with the above picture.   
De Propris et al. (1998) derived the $H$-band luminosity function of the Coma cluster galaxies, providing further support to the 
hypothesis of distinct populations: intrinsically bright red galaxies and intrinsically faint blue dwarf galaxies. 
The dwarf-to-giant ratio also appears to depend on environment. Several studies (Driver et al. 1998; Phillips et al. 1998; 
 Andreon 2001; Dahl{\' e}n et al. 2004) find that dwarf galaxies are more common in lower density environments in local clusters. 

It is important to extend these studies to higher redshifts as the evolution of the above characteristics can provide useful discrimination amongst
structure formation and evolution models. To date, however, only a handful of clusters at $z\sim1$ have been studied in 
detail (Stanford et al. 1997; Benitez et al. 1999; Rosati et al. 1999; Tanaka et al. 2000; Haines et al. 2001; van Dokkum et al. 2001; Lidman et al. 2004). 
In addition, most of these studies relied on photometric redshifts to determine cluster membership.
The evolution of the $K$-band luminosity function in the redshift range $z=0.1$-$0.9$ has been studied by De Propris et al. (1999). The
evolution of the characteristic magnitude $K^*$ of the galaxies, is found to be consistent with a passively evolving population of
galaxies formed at $z_f>2$. The data are not deep enough, however, to constrain the faint-end slope 
which was fixed at $\alpha=-0.9$ in the above study (and adopted from the value derived in the $H$-band for the Coma cluster). 
 Nakata et al. (2001) extended the study to $z=1.2$ by deriving the $K$ band
luminosity function for the cluster around the radio galaxy 3C324. 
They also chose to fix the faint-end slope at  $\alpha=-0.9$
and found a result consistent with what is expected for a passively evolving
stellar population formed at $z_f>2$. Toft et al. (2004) studied the $K$-band LF of galaxy 
cluster RDCS J1252.9-2927 at $z=1.237$. They found $\Delta M_z^* =1.3\pm0.5$ mag of evolution 
and a similar shape for the faint-end tail.
The main difficulties in studying cluster galaxy populations at $z\sim1$ are, of course,
the fainter apparent magnitude of the distant cluster galaxies and contamination by non-cluster member galaxies. 

With the advent of the Advanced Camera Surveys (ACS) on board the {\it Hubble Space Telescope (HST)},
we now have a chance to step forward. As a part of Guaranteed-Time-Observations (GTO), 
we have observed the galaxy cluster, MS1054-0321 at $z=0.83$. 
A large companion spectroscopic campaign yielded redshifts for 143 cluster members, 
eliminating the uncertainties associated with photometric redshift estimation or statistical background 
subtraction (Driver et al. 1994; Bernstein et al. 1995; Valotto et al. 2001). 
The superb sensitivity and angular resolution of the ACS allows us to classify galaxies morphologically to the depth of $i_{775}=24$ mag. 

We present in this paper our ACS-based LF and its dependence on clustocentric position, environment, and galaxy properties for  
the galaxy cluster MS1054-0321 at $z=0.83$. 
The paper is organized as follows: In Section~\ref{data}, we describe the observations;
in Section~\ref{analysis}, we describe the details of the analysis; in Section~\ref{results}, we present the results;
in Section~\ref{discussion}, we discuss the physical implications of our results; and in Section~\ref{conclusion},
we summarize our work and findings. Unless otherwise stated, we adopt the best-fit WMAP cosmology:
 $(h,\Omega_m,\Omega_L) = (0.71,0.27,0.73)$ (Bennett et al. 2003), giving a scale of 7.62 kpc per arcsec at $z{\,=\,}0.83$.

\section{Data}\label{data}

\subsection{MS1054-0321}

With a redshift of $z=0.831$, MS1054-0321 is the most distant cluster in the Extended 
Medium Sensitivity Survey (EMSS) x-ray selected cluster sample (Gioia et al. 1990). 
It has a rest-frame x-ray luminosity of $L_X$(2-10keV)=2.2$\times 10^{45} h_{50}^{-1}$ 
erg s$^{-1}$(Donahue et al. 1998), which makes it one of the brightest x-ray clusters known.
  Recent observations with XMM (Gioia et al. 2004) find that
 the X-ray temperature is 7.2$^{+0.7}_{-0.6}$ keV, in excellent agreement with Tozzi et al. (2003), Vikhlinin et al. (2002) and Donahue (2004; priv. comm.).
MS1054-0321 has an Abell richness class of 3 and a line-of-sight 
velocity dispersion of $\sigma=1170\pm 150$ km s$^{-1}$ (Tran et al. 1999). 
The weak-lensing mass is estimated to be 9.89$\pm$0.36 $\times 10^14 M_{\odot}$ (Jee et al. 2005).

A significant fraction of cluster galaxies, 17\%, are classified as 'merger/peculiar' on the basis of 
double nuclei (separation $<10$ kpc), tidal tails, and distorted morphologies (van Dokkum et al. 1999). 
Interestingly, many of the merging galaxies are red, bulge-dominated galaxies with no detected nebular line emission. 
The fraction of blue galaxies in the cluster, calculated in the same way as the original work by Butcher \& 
Oemler (1978,1984), is 0.22$\pm$0.05 (van Dokkum et al. 2000), comparable with the mean value 
determined for clusters at $0.3<z<0.5$ (e.g., Goto et al. 2003a). 

\subsection{HST ACS and Keck LRIS Observations}

We have observed MS1054-0321 in the $F606W$, $F775W$ and $F850LP$ bandpasses 
(hereafter $V_{606}$, $i_{775}$, and $z_{850}$, respectively)
with the  Wide Field Channel (WFC) of ACS as a part of the GTO program 9290.  
A mosaic observation was constructed using a $2{\times}2$ WFC pointing pattern, with 1, 2, and 2 orbits
of integration in $V_{606}$, $i_{775}$ and $z_{850}$ bands, respectively, at each 
of the four pointings.  There was nearly 1 arcmin of overlap between 
pointings; thus, the core of cluster was imaged for a total of 8 orbits in $i_{775}$
and $z_{850}$. The field of view of the mosaic is 36.51 square arcminutes covering 
an approximately square geometry with a projected proper spatial
dimension of 2.7 Mpc at $z=0.83$.
The data were processed with the ``Apsis'' pipeline described by Blakeslee et al. (2003). 
We use the SExtractor {\tt MAG\_AUTO} value output by the pipeline for our photometric measure. 
This magnitude is intended to give a precise estimate of total magnitudes for
galaxies using Kron's first moment algorithm (Blakeslee et al. 2003).
We calibrate our photometry to the AB system using photometric zero points
of 26.48 ($V_{606}$), 25.65 ($i_{775}$) and 24.86 ($z_{850}$).
These are uncertain at the $\sim\,$0.02 mag level, which has no effect on our conclusions.
We adopt a Galactic reddening for this field of
$A_V=0.0945,A_i=0.0714$, and $A_z=0.0521$ mag based on the Schlegel et al. (1998) dust maps.

Multi-object spectroscopy of the galaxies in region was carried out on the Keck Telescope using the Low Resolution Imaging Spectrograph (Oke et al. 1995).
This survey has yielded redshifts for over 300 objects, with 143 of these being confirmed as cluster members (van Dokkum et al. 2000; Tran et al. in prep.). Spectroscopic redshifts are measured for 76\% of the galaxies brighter than $i_{775} = 22.0$  (See Section~\ref{correction}). 
The availability of this extensive spectroscopic database provides a significant advantage 
in studying the galaxy cluster MS1054-0321. To obtain a large number of spectroscopic memberships in 
galaxy clusters at $z\sim 1$, a large amount of telescope time is usually required, and thus, most of the 
previous studies had to rely on the photometric redshift estimation or the statistical background subtraction, 
both of which were the main source of bias in determining the slope of the LF at the 
faint end (Driver et al. 1994; Bernstein et al. 1995; Valotto et al. 2001).
We exclusively use the spectroscopically confirmed cluster members to construct the LF in MS1054-0321.

\section{Analysis}\label{analysis}

\subsection{Cluster Membership Determination}

We regard the 143 galaxies with $0.81<z<0.85$ as members of MS1054-0321. This corresponds to a rest-frame spread of $\pm3300$ km s$^{-1}$ about the mean cluster redshift.  
After correcting to the cluster restframe and subtracting the contribution to the observed velocity dispersion from the redshift measurement errors ($\sim 200$ km s$^{-1}$), we find a line-of-sight cluster velocity dispersion of 1131$\pm$5 km s$^{-1}$ (measured as a sigma of biweighted mean with the jackknife error estimate). 
The derived line-of-sight velocity dispersion is consistent with previously published numbers (e.g., Tran et al. 1999; Goto et al. 2004b).

\subsection{Correction for Incompleteness in Spectroscopic Sample}\label{correction}

\begin{figure*}
\begin{center}
\includegraphics[scale=0.8]{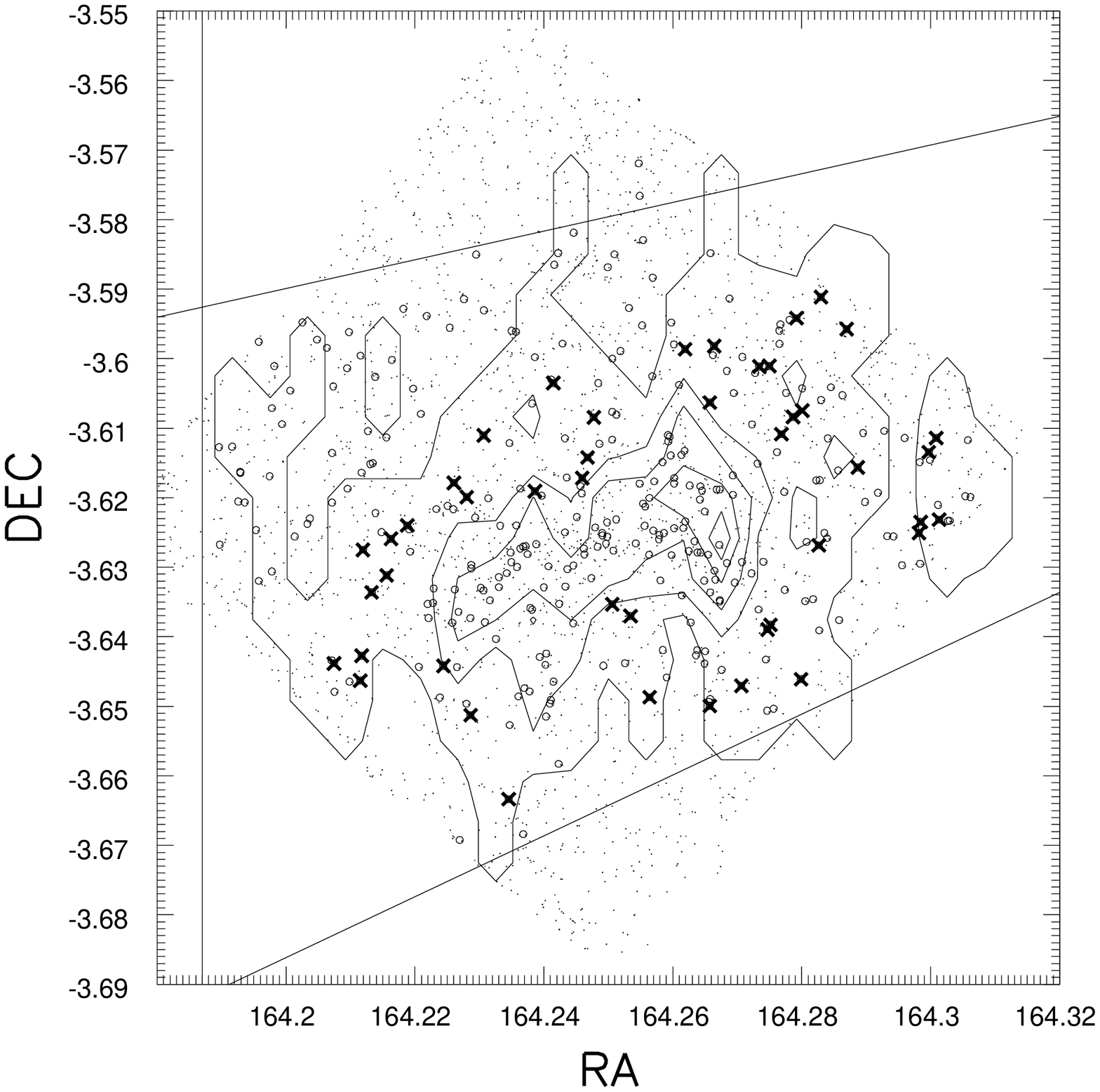}
\end{center}
\caption{
 The sky distribution of all extended objects brighter than $i_{775}=25$. The large circles are galaxies with spectroscopic redshifts. The small dots are extended objects in the imaging data with $i_{775}<25$. The contours are based on the number density of spectroscopically confirmed cluster members. The crosses are cluster members with the local galaxy density of $10<\Sigma<40$ Mpc$^{-2}$.
}\label{fig:radec}
\end{figure*}

\begin{figure}
\begin{center}
\includegraphics[scale=0.8]{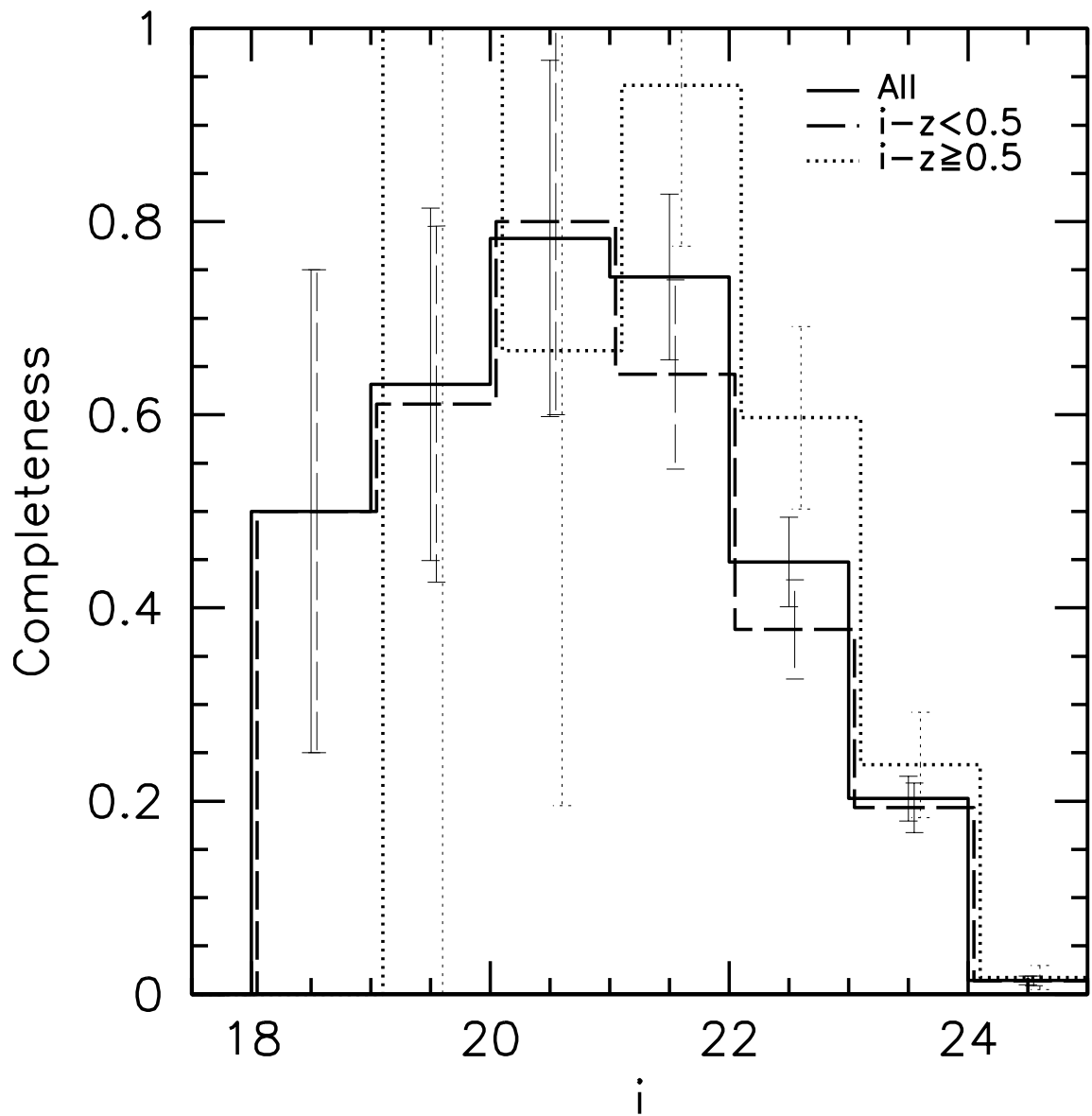}
\end{center}
\caption{Completeness in obtaining spectroscopic redshift as a function of apparent magnitude in $i_{775}$. The number ratio of galaxies with spectroscopic redshift with respect to all the galaxies in the spectroscopic area of the imaging data is plotted against $i_{775}$ mag. The error bars assume Poisson statistics. The bin size is 1.0 mag. The dashed and dotted lines are for blue ($i_{775}-z_{850}<0.5$) and red ($i_{775}-z_{850}\geq 0.5$) samples, respectively.
}\label{fig:incompleteness}
\end{figure}

Incompleteness in measuring spectroscopic redshifts (or targeting galaxies for observation) as a function of apparent magnitude must be compensated for in order to extract unbiased results. Fig.~\ref{fig:radec} shows the sky distribution of objects with spectroscopy (circles) and all extended objects in the imaging data (small dots). Since we have only performed spectroscopic observations in the region defined by the solid lines, we use objects within this region to define and quantify the incompleteness (see van Dokkum et al. 1999,2000; Postman et al. 2004; Tran et al. in prep.  for more details of the spectroscopic observations). Point sources are excluded from incompleteness calculations using SExtractor's stellarity index ({\tt CLASS\_STAR} $>0.5$), which assigns a numerical value close to 1 if the object is a point source and values closer to 0 if it is an extended object. A slight change in the criterion of the separation between extended and point sources ({\it e.g.,} using a {\tt CLASS\_STAR} value of 0.8 instead of 0.5 as the threshold) does not affect our results.  In Fig.~\ref{fig:incompleteness}, we show the fraction of galaxies that have a spectroscopic redshift among all galaxies in the imaging data as a function of apparent $i_{775}$ magnitude. As can be seen in Fig.~\ref{fig:incompleteness}, the completeness is almost independent of magnitude in the range of $18.0<i_{775}<22.0$, indicating that the spectroscopic survey is not strongly biased and has a high degree of completeness. Regardless, when we measure a LF in the following sections, we correct for the incompleteness by weighting the observed counts by the inverse of the completeness.
  
In the following analyses, we divide our cluster galaxy sample using various criteria in morphology,
environment, and color. When measuring LFs of these subsamples, the incompleteness correction is re-computed 
using each subsample of galaxies, and then applied to the LFs of each subsample accordingly. 
As an example, we plot the spectroscopic completeness for blue ($i_{775}-z_{850}<0.5$) and red ($i_{775}-z_{850}\geq 0.5$) subsamples in dashed and 
dotted lines in Fig.\ref{fig:incompleteness}. It can be seen that the completeness is not strongly dependent on $i_{775}-z_{850}$ color.
When we construct a LF in different passbands ($V_{606}$, $i_{775}$ and $z_{850}$), we correct the counts using
the incompleteness as a function of magnitude in that passband. 

\subsection{Quantifying the Environment of Galaxies}\label{environment}

Because different physical mechanism that act upon galaxies in dense environments have efficiencies that vary as a function of density or position, we quantify the LF as a function of these variables in a hope of understanding the underlying processes (see a review in Treu et al. 2003).
 
\subsubsection{ Local Galaxy Density}\label{density}

 For each galaxy, we measure a projected (angular) distance to the 5th nearest galaxy using the cluster members ($0.81<z<0.85$).  The number of galaxies ($N=5$) within the distance is divided by the circular surface area with the radius of the distance to the 5th nearest galaxy to obtain the local galaxy density, $\Sigma$ (Mpc$^{-2}$).  When the projected area touches the boundary of the data (Fig.~\ref{fig:radec}), we calculate the density by dividing the area within the data region.
This parameterization is similar to the one used in previous work (e.g., Goto et al. 2003b,c; Tanaka et al. 2004).
 Although our local galaxy density is a two-dimensional surface density, the contamination from both foreground and background galaxies is zero
 because we are using spectroscopic information. 
 However, because we do not have spectroscopic redshifts for all the members, our measured density is an underestimate. We use this density only for internal comparisons. 
 We have chosen the 5th nearest to measure local galaxy density, but we also have confirmed that using up to the 10th nearest
 neighbor does not change the main results described below.

\subsubsection{Clustocentric Distance}

\begin{figure}
\begin{center}
\includegraphics[scale=0.8]{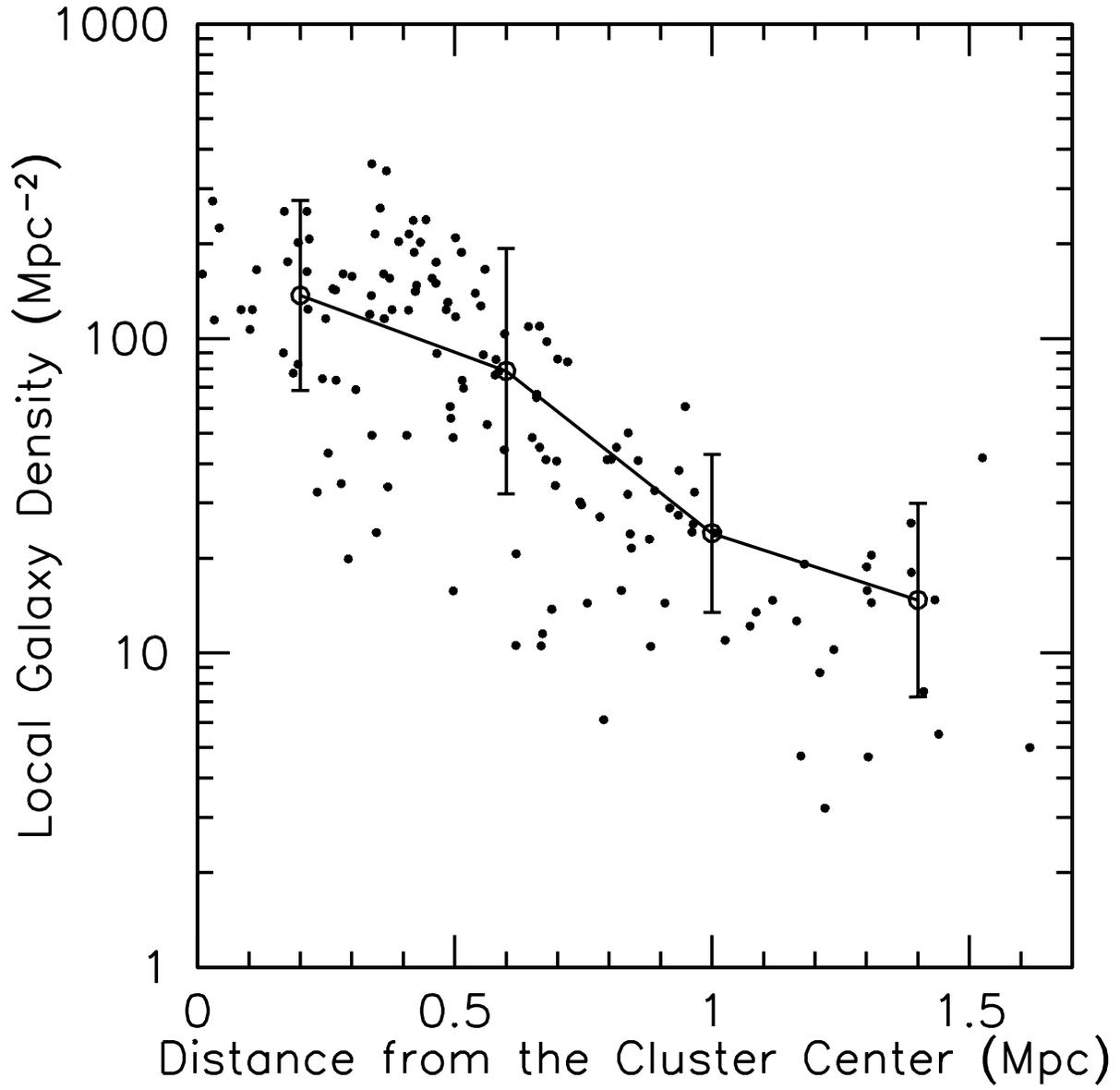}
\end{center}
\caption{ Local galaxy density is plotted against the clustocentric distance.
}\label{fig:distance_density}
\end{figure}

To assess dependencies of the LF on position within the cluster,
we measure the projected distance to the cluster center, $R$, for all spectroscopically confirmed member galaxies. 
We use the position of the brightest cluster galaxy (J2000 RA: 10h 56m 59.99s, Dec: -03d 37m 36.1s) as the center of the cluster. 
Using the centroid of the x-ray surface brightness distribution (Neumann \& Arnaud 2000) does not significantly change the results.
The virial radius of the cluster is known to be $\sim$1 Mpc (Tran et al. 1999). 
In Fig.\ref{fig:distance_density}, we compare the clustocentric distance with the local galaxy density. 
The solid lines connect medians. The error bars are based on the rms in each bin. 
As expected, there is a fairly good correlation between these two parameters. 
Note that clustocentric distance of $R=$0.9 Mpc corresponds to 30-40 Mpc$^{-2}$ in local galaxy density.

\subsection{Luminosity Function Measurements}

We use a non-linear least squares method to determine the best-fit Schechter (1976) LF parameters.  The LF normalization ($\phi^*$), faint-end slope ($\alpha$), and characteristic magnitude (M$^*$) are fit simultaneously. The input data for the fitting procedure is a binned, apparent magnitude distribution for the spectroscopically confirmed cluster members, where the number of galaxies in each bin has been corrected for incompleteness in the spectrscopic survey. We use a magnitude bin size of 0.5 mag. The values of the best fit LF parameters do not strongly depend on our choice of the bin size in the range $0.25 < \delta m < 0.75$. 
The incompleteness correction is determined empirically by dividing the number of objects with measured redshifts in a given apparent magnitude bin by the total number of galaxies in that same bin. The total number of galaxies in a bin is taken from our SExtractor catalog. The corrected number counts as a function of apparent magnitude are then converted to the corresponding absolute magnitude using a distance modulus of $m-M=42.86$. We do not apply k-corrections for the internal MS1054 LF comparisons. However, whenever we compare our results to those from other studies, a k-correction is applied and the details of how the k-correction is determined are given in Section \ref{comaprion_literature}. Finally, we divide the corrected number counts by the effective area of the survey (in units of projected Mpc$^2$; see Fig.\ref{fig:radec}).  All fits are weighted by the inverse square of the error in the galaxy number in each bin and these errors are determined from Poisson statistics using the original (uncorrected) number counts. The best-fit Schechter parameters for all samples are summarized in Table \ref{tab:lf_parameter}.

\section{Results}\label{results}

\subsection{Luminosity Function of Galaxy Cluster MS1054-0321}

With the spectroscopic membership information and parameters to describe the environment of each galaxy in hand, it is straight-forward to construct a LF of the member galaxies. It is desirable to normalize the absolute magnitude to $h=1$ for external comparison purposes, and we thus adopt the distance modulus of $m-M=42.86$ at $z=0.83$.  In the upper panel of Fig.~\ref{fig:lf_3mag}, we show the LFs of the spectroscopic members. The incompleteness in the spectroscopic sample is corrected for by using the prescriptions described in Section~\ref{correction}.  The error bars assume Poisson statistics. The overall normalization is shown in Mpc$^{-2}$, obtained by dividing the corrected counts in each bin by the surface area of the survey.  The dotted, dashed, and solid lines are LFs in $V_{606},i_{775},$ and $z_{850}$ filters, respectively.  The best-fit Schechter functions (Schechter 1976) are shown in the lower panel of Fig.~\ref{fig:lf_3mag}.  Fits to the luminosity function are performed using a non-liner least square method applied to the corrected binned number counts, where each bin is weighted by its error.  The best-fit parameters are ($M^*_V,\alpha$) = ($-$19.59$\pm$0.24, $-$1.35$\pm$0.13) , ($M^*_i,\alpha$) = ($-$20.83$\pm$0.16, $-$0.82$\pm$0.10), and ($M^*_z,\alpha$) = ($-$21.47$\pm$0.29, $-$0.87$\pm$0.15). These results are also summarized in Table~\ref{tab:lf_parameter}.

Our LF parameters are, of course, less accurate than those derived from
composite LFs from many galaxy clusters (e.g., Goto et al. 2002b). 
Nevertheless, characteristic features of cluster LFs can be clearly recognized.
In the literature, LFs in bluer bands tend to have steeper faint-end slope, 
which is clearly seen in Fig.~\ref{fig:lf_3mag}. At $z=0.83$, $V_{606}$ filter 
corresponds to the rest-frame $UV$ light. The difference between the LF in  $V_{606}$ and 
the LFs in $i_{775},z_{850}$ is a nice distinction between star-forming and old-passive  galaxies. 
We compare our LFs with those in the literature in Section~\ref{discussion}.
 
\subsection{Luminosity Function as a Function of Clustocentric Distance}

We begin to explore the dependence of the LF on environment by looking for differences in
the galaxy luminosity distribution as a function  of clustocentric distance.
In the upper panel of Fig.~\ref{fig:lf_distance}, we show the LF in the inner ($R < 0.9$ Mpc) and outer ($R \geq$ 0.9 Mpc)  regions of MS1054-0321.
The threshold of 0.9 Mpc is chosen to correspond to the distance where galaxy properties start to undergo
significant changes inwards towards the core.
The cluster core region with $R< 0.9$ Mpc  has a higher density of bright galaxies 
and a flatter faint-end slope compared to the LF at $R\geq 0.9$ Mpc.
We fit a Schechter function to the observed data and show the results in the lower panel of  
Fig.~\ref{fig:lf_distance} along with the errors on each data point. 
These error bars assume Poisson statistics. 
The best-fit Schechter parameters are ($M^*_i$,$\alpha$) = ($-$20.92$\pm$0.76, 
$-$1.28$\pm$0.35), ($-$20.77$\pm$0.24, $-$0.64$\pm$0.17) for galaxies with 
$R\geq 0.9$ and $R< 0.9$, respectively (see also Table~\ref{tab:lf_parameter}). 
A steeper faint-end slope is detected for galaxies in the outer part ($R\geq 0.9$ Mpc) 
as compared with the faint-end slope for galaxies in the
inner cluster region.  These trends are in general agreement with the findings in the 
composite LFs of galaxy clusters in the literature (e.g., Goto et al. 2002b). The difference in LFs in different environments 
suggests that the environment indeed plays a role in determining the luminosities of cluster galaxies.  

\subsection{Luminosity Function as a Function of Local Galaxy Density}

In the upper panel of Fig.~\ref{fig:lf_density}, we plot the LF of the cluster partitioned by the local galaxy density of 30 Mpc$^{-2}$. The solid line is for galaxies in regions with $\Sigma\geq$ 30 Mpc$^{-2}$ and the dashed line is for galaxies in regions with $\Sigma <$ 30 Mpc$^{-2}$. 
Again, the criterion of $\Sigma=$ 30 Mpc$^{-2}$ is chosen to correspond to the environment where galaxy properties start to exhibit
substantial change relative to the field galaxy population (see Section~\ref{sec:gdr}).
The LF of galaxies in dense regions has a flatter faint-end slope than in sparse environments, consistent with the results obtained as a function
of clustocentric radius. Note that we do not normalize the LFs in Fig.~\ref{fig:lf_density} because 
the sampling volume of each galaxy is different as a consequence
of the way the local galaxy density is measured (Section~\ref{density}). 
In the lower panel, we fit a Schechter function to each LF.
The best-fit Schechter parameters are ($M^*_i$,$\alpha$) = ($-$21.43$\pm$0.71, $-$1.42$\pm$0.25), 
($-$20.63$\pm$0.15, $-$0.53$\pm$0.12) for galaxies with $\Sigma < 30$ Mpc$^{-2}$ and 
$\Sigma \geq 30$ Mpc$^{-2}$, respectively. These results  are also given in Table~\ref{tab:lf_parameter}. 
Although the errors on the parameters are substantial, the parameter values are consistent with the 
interpretation that the galaxy LF in dense regions has a flatter faint-end slope. 

\subsection{Luminosity Function as a Function of Color}

\begin{figure}
\begin{center}
\includegraphics[scale=0.8]{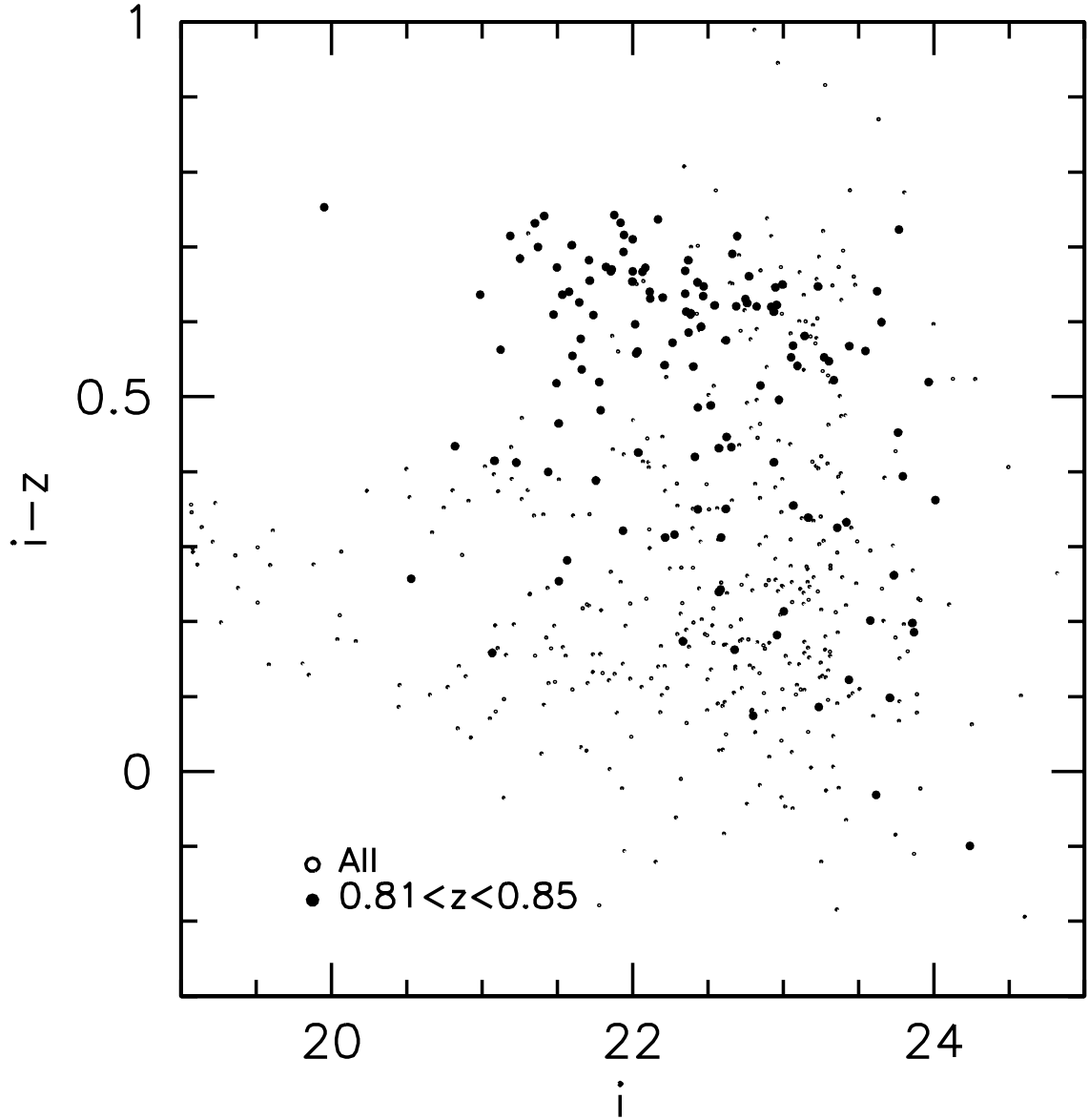}
\end{center}
\caption{ A color-magnitude diagram. The  $i_{775}-z_{850}$ color is plotted against apparent $i_{775}$ magnitude for all objects (the black dots) and for cluster members with spectroscopic redshift (the large circles). 
}\label{fig:cmd}
\end{figure}

\begin{figure}
\begin{center}
\includegraphics[scale=0.4]{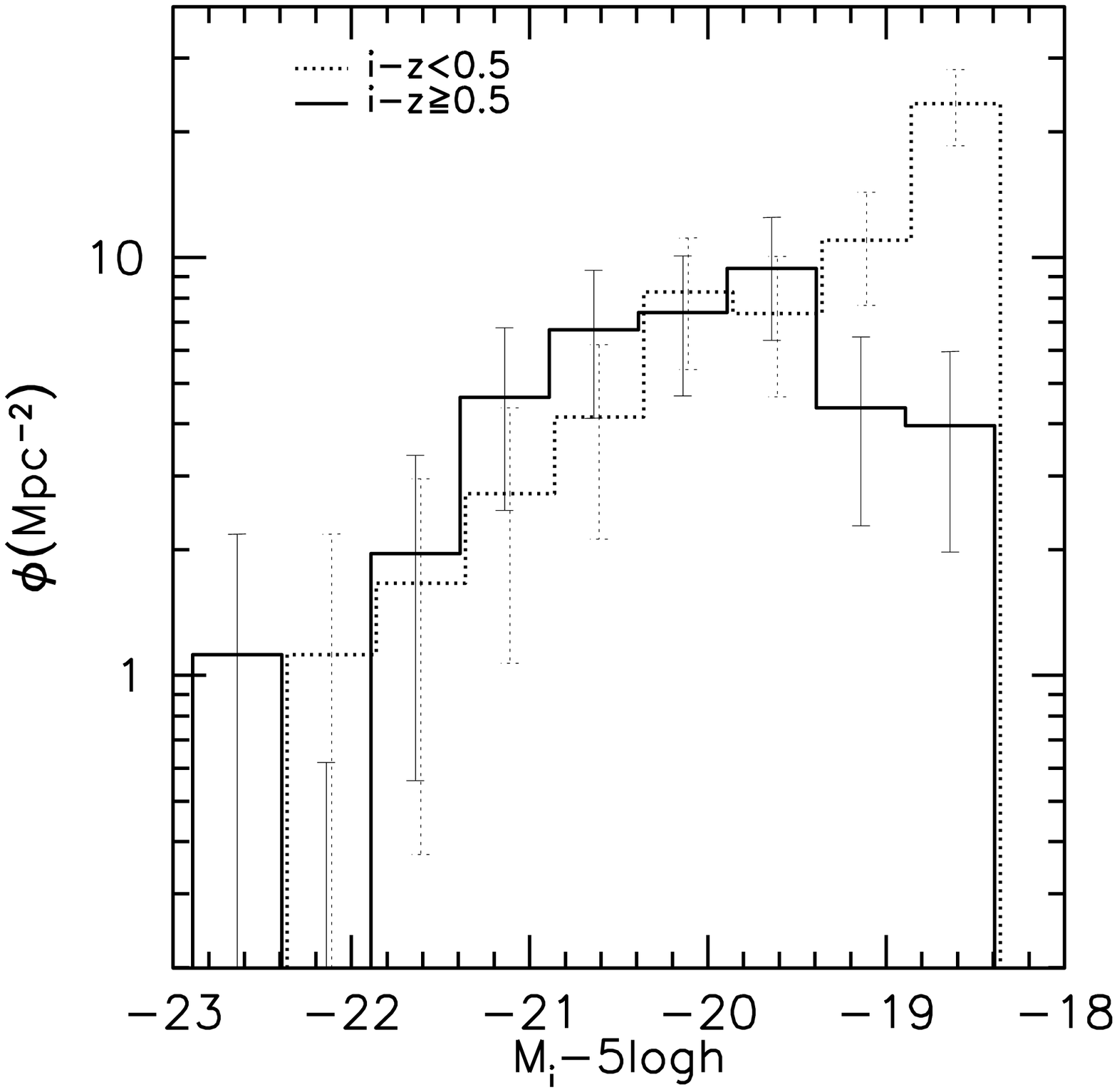}
\includegraphics[scale=0.9]{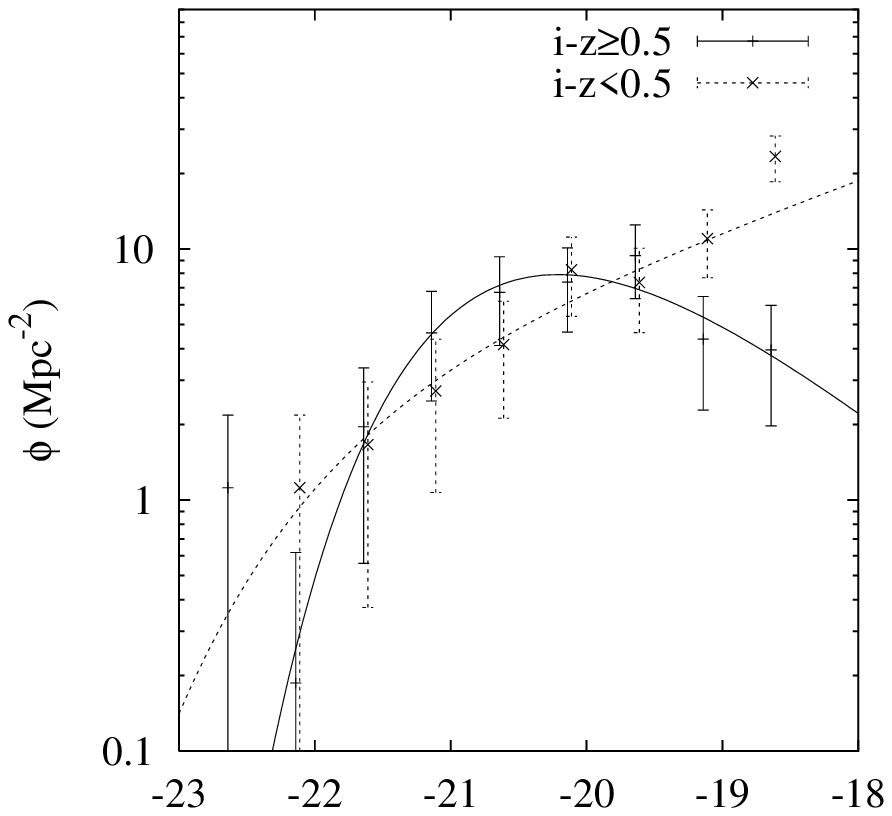}
\end{center}
\caption{ LFs divided at $i_{775}-z_{850}$ color of 0.5. The corrections for the incompleteness of spectroscopic observation is corrected for each subsample of galaxies.
}\label{fig:lf_iz}
\end{figure}

The $i_{775}$ filter is particularly sensitive to different stellar populations at $z=0.83$ because it
includes the 4000\AA\ break. 
In particular, the $i_{775}-z_{850}$ color does a reasonable job of distinguishing blue, star-forming galaxies from 
red, passive galaxies in the color-magnitude relation (Homeier et al. in prep.).
Fig.~\ref{fig:cmd} shows the color-magnitude relation for the known cluster members (filled circles) 
and for all extended objects in the field (small dots). The $i_{775}-z_{850}=0.5$ separates well the galaxies in the 
red-sequence from the other galaxies. A detailed analysis of ACS-based color-magnitude relation in MS1054-0321 
will be discussed in Blakeslee et al. 2004.
The upper panel of Fig.~\ref{fig:lf_iz} shows the LFs of galaxies with ($i_{775}-z_{850}<0.5$) and ($i_{775}-z_{850}\geq0.5$) by 
the dashed and solid lines, respectively. 
The corresponding best-fit Schechter functions are shown in the lower panel of Fig.~\ref{fig:lf_iz} with best-fit parameters of ($M^*_i$,$\alpha$) =
($-$20.11$\pm$0.22,$-$0.09$\pm$0.27), ($-$21.92$\pm$0.86, $-$1.48$\pm$0.23) for red and blue samples, respectively. 
One potentially interesting difference is that the LF of red galaxies shows a deficit at the faint end $M^*_i>-19$. 
Similar deficits of faint red galaxies in high redshift galaxy samples have been reported in the literature (e.g., Kajisawa et al. 2000; Nakata et al. 2001; Kodama et al. 2004; De Lucia et al. 2004). Such a deficit may signal the presence of galaxy population that has yet to evolve into a faint, red phase. 
We discuss this hypothesis and its implication further in Section ~\ref{discussion}.

\subsection{Luminosity Function as a Function of Galaxy Morphology}

In addition to providing superb quality photometry, a key advantage provided by the high angular resolution provided by HST and the ACS, in particular, is the ability to classify galaxy morphology out to redshifts as high as $z=0.83$. One of us (M.P.) has visually classified the morphology of MS1054-0321 galaxies with $i_{775}<24$ (see Postman et al. 2004 for further details).  We use these classifications here to examine how the luminosity distribution depends on morphology.  In Fig.~\ref{fig:lf_morph}, we present a LF of early-type galaxies (E+S0 or $T \le 0$) and that of late-type galaxies (Sa or later; $T > 0$).
Our best-fit Schechter parameters are ($M^*_i$,$\alpha$) =  ($-$20.76$\pm$0.12, $-$0.54$\pm$0.13), ($-$20.79$\pm$0.79, $-$0.84$\pm$0.67) for early- and late-type galaxies, respectively. The best-fit Schechter functions are shown in the lower panel of the figure.  To highlight some subtle differences, we plot the ratio of morphologically classified galaxies as a function of absolute magnitude in Fig.~\ref{fig:morph_ratio} with slightly finer bins.  There may be a gradual decline of early-type fractions toward fainter absolute magnitudes,
 suggesting that the cluster LF is dominated by early-type galaxies.  Note that at the bright-end the fraction of early-type galaxies is somewhat higher than previous estimates (44\%; van Dokkum et al. 2000). This is due to a difference in morphological classification process as we attempt to assign a Hubble type to galaxies involved in mergers, if at all possible, to facilitate comparisons with work at lower redshift (see Postman et al. 2004 for more details in morphological classification).

\subsection{Giant-to-Dwarf Ratio}\label{sec:gdr}

\begin{figure}
\begin{center}
\includegraphics[scale=0.8]{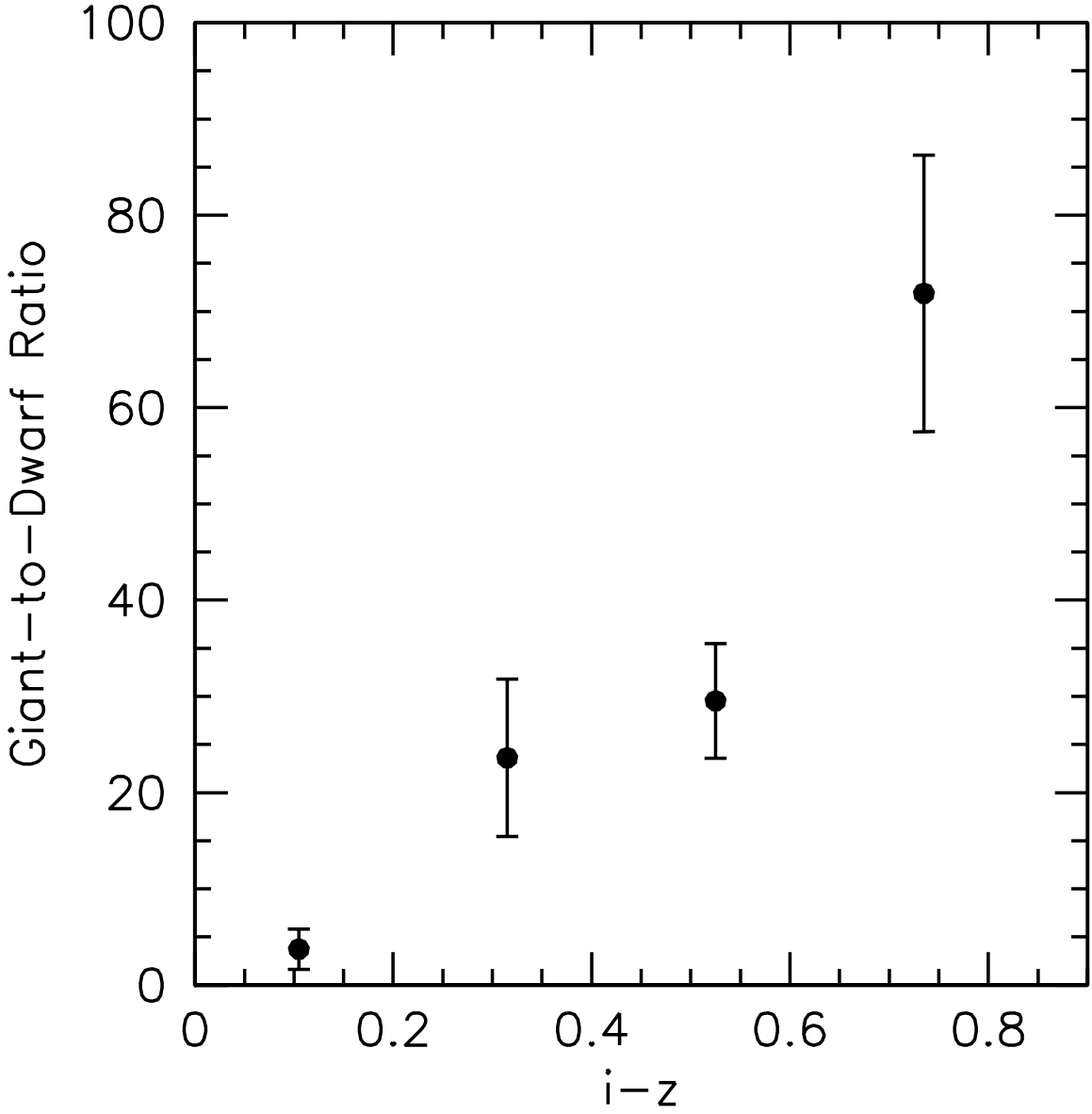}
\end{center}
\caption{Giant-to-Dwarf ratio as a function of $i_{775}-z_{850}$ color.
}\label{fig:gdr_iz}
\end{figure}

\begin{figure}
\begin{center}
\includegraphics[scale=0.8]{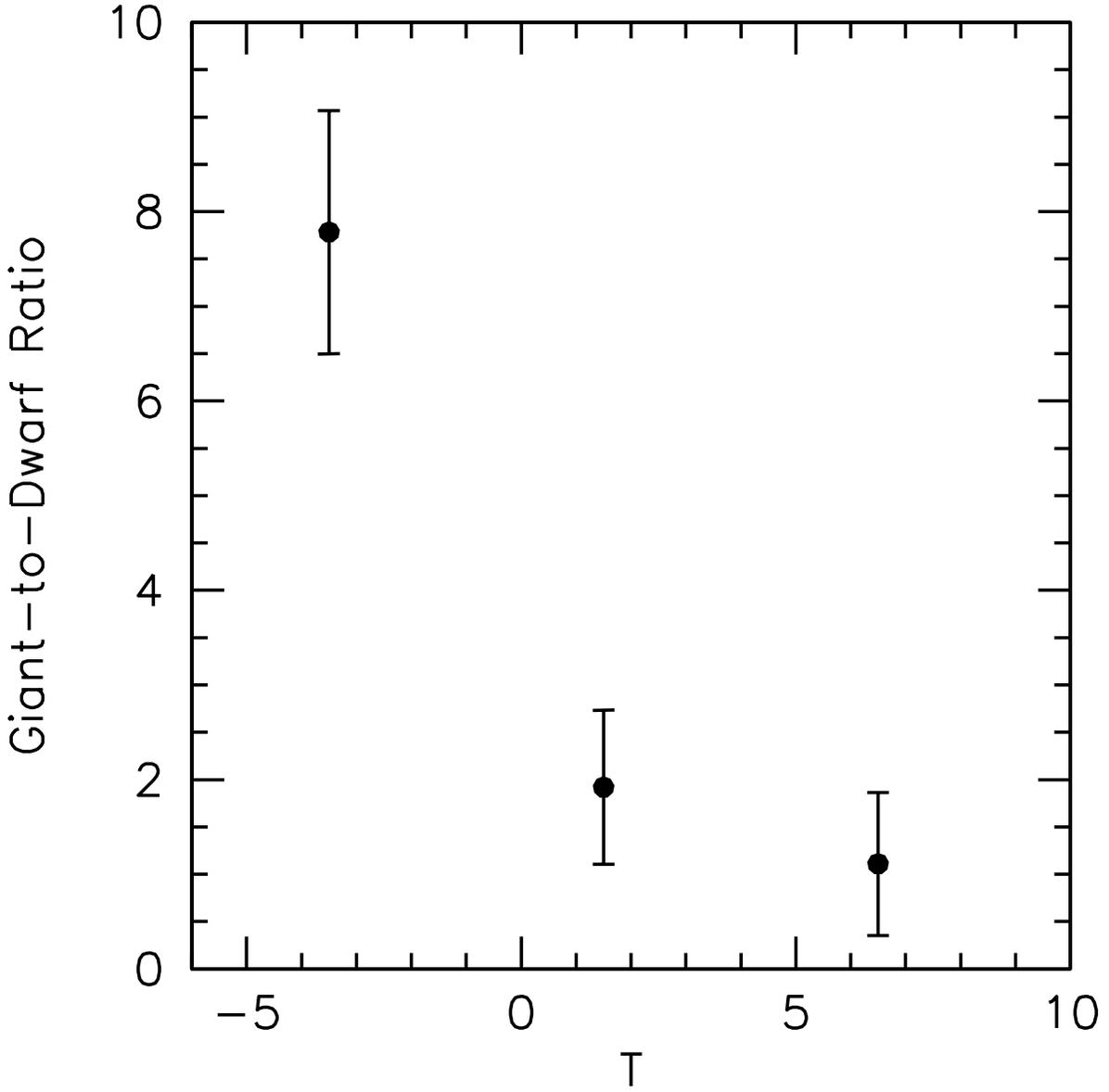}
\end{center}
\caption{Giant-to-Dwarf ratio as a function of morphology.
}\label{fig:gdr_morph}
\end{figure}

In this section, we investigate the giant-to-dwarf ratio (GDR) as functions of color, morphology, clustocentric radius, and local projected density.  
We calculate the GDR as a number ratio of giant galaxies ($M_i<-19.5$) to dwarf galaxies ($-19.5\leq M_i< -18.0$) using the expression: 
 \begin{equation} 
    GDR=\frac{N(M_i<-19.5)}{N(-19.5\leq M_i<-18.0)}.\label{eq_gdr}
 \end{equation}
The criterion between dwarf and giant at $M_i=-19.5$ is chosen to correspond approximately to the start of the deviation between the faint ends of the color-selected LFs in Fig.~\ref{fig:lf_iz}.  Due to the large distance to the cluster, we cannot sample the dwarf population as faint as was probed in previous studies ($\sim-16.5$ mag; e.g., Ferguson \& Sandage 1991; Secker \& Harris 1996).  Therefore, we take the inverse of the commonly used dwarf-to-giant ratio in order to magnify the subtle change in the parameter.  In Fig.\ref{fig:gdr_iz}, we show the GDR as a function of $i_{775}-z_{850}$ color.  The trend seen in Fig.~\ref{fig:lf_iz} is much clearer here in the sense that the GDR is much higher for red galaxies, and it gradually declines as the galaxy color becomes bluer.  In Fig.~\ref{fig:gdr_morph}, we show the GDR as a function of morphological type.  Confirming the trend seen in Fig.~\ref{fig:lf_morph}, the GDR is higher for early-type ($T \le 0$) galaxies.  These two trends are consistent with the hypothesis that the bright end of the cluster LF is dominated by red or early-type galaxies and the faint end is dominated by blue or late-type galaxies.

In previous sections, we have shown that galaxies in the cluster core (or the densest regions) show different characteristics from galaxies in the cluster outskirts (or the sparsest regions).  An important question to ask now is where (in what environment) these changes happen. In Fig.~\ref{fig:gdr_radius}, we show the GDR as a function of clustocentric distance.  The GDR is higher for galaxies in the inner regions of the cluster, consistent with the results shown in Fig.~\ref{fig:lf_distance}.  In addition, Fig.~\ref{fig:gdr_radius} shows that the GDR rises steeply at $R< 0.9$ Mpc, and flattens outside this radius.  Similarly in Fig.~\ref{fig:gdr_density}, we show the GDR as a function of the local galaxy density ($\Sigma$) defined in Section~\ref{density}. The same trend can be seen here. The GDR rises with increasing local galaxy density.  At $\Sigma<30$ Mpc$^{-2}$, the GDR is consistent with a constant.  If MS1054-0321 were roughly spherical, the de-projected rise in the GDR at small radii would be even steeper. However, MS1054-0321 is far from spherical - as evidenced by the distributions of  the X-ray gas (Donahue et al. 1998; Neumann \& Arnaud 2000; Jeltema et al. 2001), the galaxies and the inferred mass (Luppino \& Kaiser 1997; van Dokkum et al. 1999,2000; Hoekstra et al. 2000; Jee et al. 2005).

\section{Discussion}\label{discussion}

\subsection{Comparison with Cluster Luminosity Functions in the Literature}\label{comaprion_literature}

We begin our discussion by comparing the derived LF of MS1054-0321 with that for cluster galaxies in the local universe. 
Since the majority of previous work has measured the LFs in the rest-frame $B$-band, we convert 
our $M^*_i$ to the rest-frame $B$-band for a fair comparison. The conversion is reliable since the shift between the sampled rest-frame at
$z=0.83$ in the $i_{775}$ filter is quite close to the $z=0$ $B-$band spectral coverage.  
As most of the cluster galaxies have passive SEDs (Blakeslee et al. 2003), we use an
elliptical galaxy SED template from Ben{\'{\i}}tez et al. (2004)  in order to compute the transformation to the $B-$band.
The transformed characteristic magnitude for MS1054-0321 in the rest-frame $B-$band is $M_B^*=-20.35$ ($h=1$). 
While not all cluster galaxies have the SED of the elliptical galaxy, the uncertainty introduced by using an elliptical SED for the
rest-frame conversion is small because (i) the majority of the galaxies in clusters do exhibit elliptical SEDs
and (ii) a spiral SED (Sbc in Ben{\'{\i}}tez et al. 2004) yields a transformation that differs by only 0.126 mag, 
comparable to the statistical error.

\begin{figure}
\begin{center}
\includegraphics[scale=0.8]{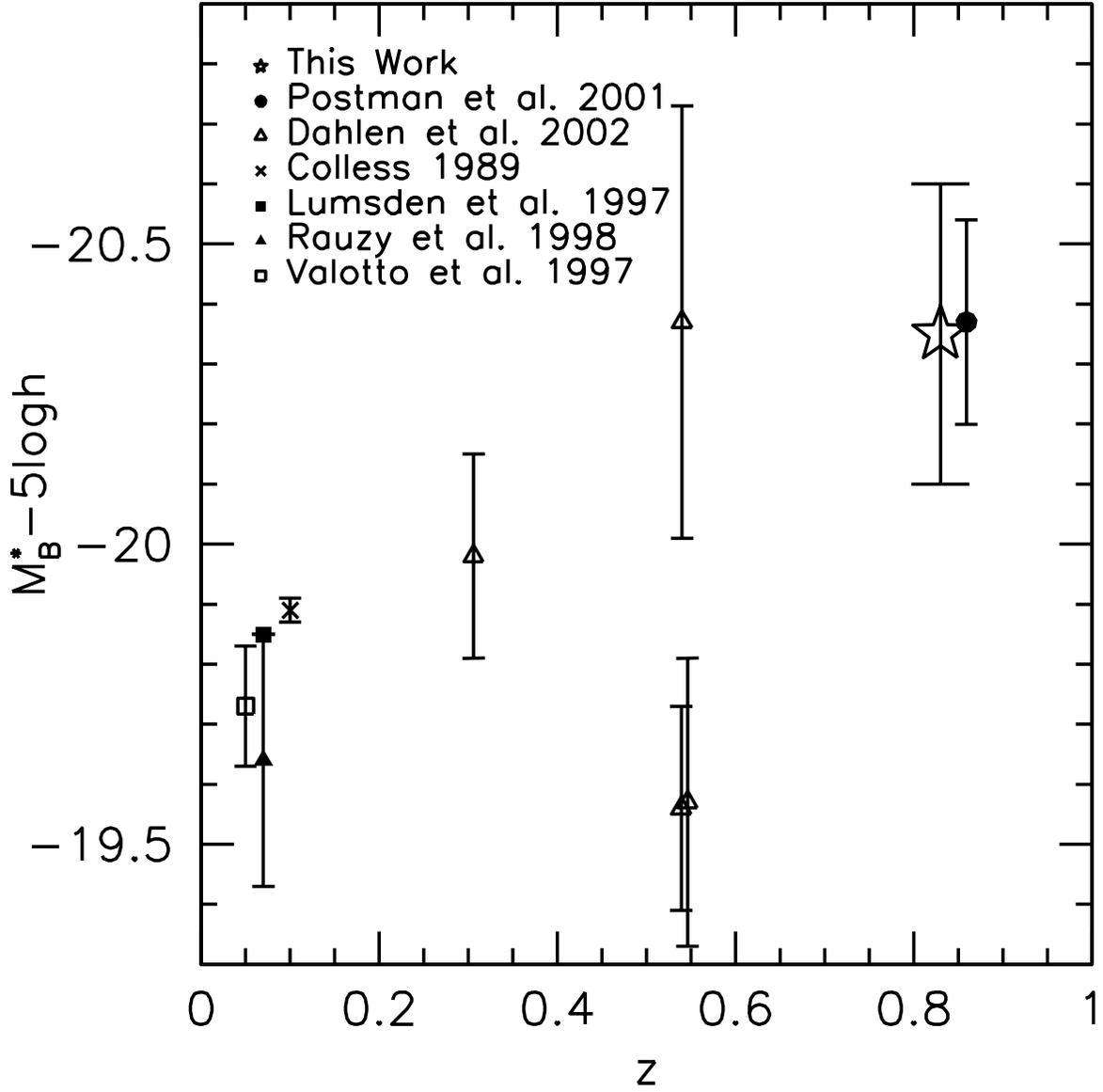}
\end{center}
\caption{Comparion to $M_B^*$ in the literature.
}\label{fig:M_B}
\end{figure}

Postman et al. (2001) measured $M_B^*=-20.37\pm 0.17$ for three clusters with a median redshift of 0.859 
(all magnitudes we quote in this section are converted to $h=1$ and are on the Vega magnitude system), 
in excellent agreement with our MS1054-0321 results.  
At low redshift, Colless (1989) measured ($M^*_{bj},\alpha$) = ($-$20.12, $-$1.24) using 14 nearby rich clusters. 
Lumsden et al. (1997) found ($M^*_{bj},\alpha$) = ($-$20.16$\pm$0.02, $-$1.22$\pm$0.04) at $z\sim 0.1$. 
Valotto et al. (1997) found ($M^*_{bj},\alpha$) = ($-$20.0$\pm$0.1, $-$1.4$\pm$0.1) at $z\sim 0.05$. 
Rauzy, Adami, \& Mazure(1998) studied 28 rich clusters with a mean redshift of 0.07, and 
found ($M^*_{bj},\alpha$) = ($-$19.91$\pm$0.21, $-$1.50$\pm$0.11).
The transformation from $B$ to $b_j$ is $B=b_j+0.28(B-V)$ (Blair \& Gilmore 1982;Cunow 1993).
Since a typical $B-V$ color is 0.96 for early-type galaxies (Fukugita, Shimasaku, \& Ichikawa 1995), 
we use $B=b_j+0.27$. Applying this conversion, we find 0.46-0.71 mag of evolution in $M_B^*$ in clusters 
between $z=0$ and 0.83. We summarize these comparisons in Fig.~\ref{fig:M_B}.  
The results from Dahl{\' e}n et al. (2002) appear to have a larger scatter because we plot each cluster 
as an individual point. However, some caution is required in the interpretation of any LF evolution 
because MS1054-0321  is extremely rich with $L_X$(2-10keV)=2.2$\times 10^{45} h_{50}^{-1}$ 
erg s$^{-1}$, and therefore, is perhaps not a progenitor of the well-studied low redshift clusters.

On the other hand, there is significant disagreement between studies on the faint-end slope, $\alpha$.  
We find a much flatter faint-end slope. As our study is limited to a single cluster some of the disagreement may
be associated with peculiarities inherent to MS1054-0321. However, the determination of the faint-end slope depends 
critically on a good background subtraction and reliable completeness corrections. Hence, some of the difference may also 
reflect the fact that our analyses are based exclusively on confirmed cluster members. If the deficit of faint red galaxies
at high redshift turns out to be real, then some of the difference between our slope and the mean faint end LF slope found
at lower redshift may also be related to this under-represented population component at $z \sim 0.8$. 
The flat slope in MS1054-0321 may thus be suggestive of a  population 
of faint red galaxies that is going to be created or accreted into clusters between $0<z<0.83$. 
A similar lack of faint galaxies at high redshift was reported by Kajisawa et al. (2000) and Nakata et al. (2001) in the 3C 324 field, 
and by De Lucia et al. (2004) in the ESO distant cluster survey at $z=0.7$ and 0.8. The physics behind the origin of such a galaxy
population remains to be explored. A hint comes from Drinkwater et al.(2003) who finds 
dwarf galaxies without any disk, possibly tidally stripped, in the Fornax cluster 
(also see Bekki et al. 2003; Mieske et al. 2004). 
However, Kodama et al. (2004) reported the lack of faint red galaxies in the high redshift 
field regions using the Subaru-XMM deep data, and thus, the creation of faint galaxies may not be cluster specific phenomena.

\subsection{Type-Specific Luminosity Functions}

The dependence of the LF on color and morphology in MS1054-0321 are shown in 
Figures~\ref{fig:lf_iz},~\ref{fig:lf_morph}, and~\ref{fig:morph_ratio}.
In the local universe,  Yagi et al. (2002) studied LFs of 10 nearby clusters and 
finds that the LF of galaxies with exponential profiles has a steeper faint-end slope than that for 
galaxies with $r^{1/4}$ profiles. Similarly, Boyce et al. (2001) studied the LF of Abell 868 ($z=0.154$) and finds 
that the presence of dwarf irregular galaxies causes the faint-end slope of the cluster LF to steepen. 
Goto et al. (2002b) used 204 galaxy clusters found in the Sloan Digital Sky Survey (Goto et al. 2002a) 
to construct composite LFs in $u,g,r,i,$ and $z$ filters. When they divide LFs by morphology and color, 
the faint-end of the LFs were steeper for late-type (or bluer) galaxies. 
We also find that the faint-end slope of the LF in MS1054-0321 is always steeper
for blue ($i_{775}-z_{850}<0.5$) galaxies. These findings are consistent with the hypothesis that the
bright end of the cluster LF is dominated by red, early-type galaxies
and the faint-end by blue, late-type galaxies. 

Considering that field galaxies at around $z\sim1$ have very different properties from their local counterpart (Kajisawa \& Yamada 2001; Cross et al. 2004), it is revealing that these typical characteristics of cluster LFs are already firmly established in a high redshift galaxy cluster at $z=0.83$, suggesting that this massive cluster most likely began to form at redshifts considerably greater than unity. A study on a large, statistically complete sample of high redshift clusters is clearly needed to assess whether the presence of similar LF features is wide spread in clusters that existed 6 -- 8 Gyr ago.

\subsection{Comparison with Field Galaxy Luminosity Functions}\label{comparison_field}

To gain a better picture of how the cluster environment impacts galaxy-scale stellar systems, we should compare our MS1054-0321 LF
results to those from low-density regions at similar redshifts. Postman, Lubin, \& Oke (2001) proposed that the evolution of 
the $B$-band LFs of field galaxies are consistent with the form $M^*(z)=M(0)-\beta \times z$, where $1<\beta<1.5$. 
Our cluster $M^*_B$ roughly agrees with the equation within the errors. Cross et al. (2004) have measured LFs of morphologically selected 
field E/S0 galaxies at $z \sim 0.75$ and find $M_B^*=-20.3\pm0.3$ with $\alpha=-0.53\pm0.17$. 
In the Subaru Deep Survey, Kashikawa et al. (2003) derives 
($M^*_B,\alpha$)=($-20.18^{+0.26}_{-0.63}$, $-1.07^{+0.01}_{-0.07}$) for field galaxies at $z\sim 1$. 
Their $M^*_B$ is slightly fainter than the value we obtained for the cluster LF. 
The $\alpha$ is much steeper than the cluster value we found.  These differences are comparable with the differences we find
when we compare LF of galaxies in the inner region of the cluster with those in the outskirts of the cluster, 
i.e., lower density regions have a fainter $M^*$ and steeper faint-end tail.  Again, these results are consistent with the 
hypothesis that cluster regions are dominated by the bright galaxies, and faint galaxies are more numerous in the field regions.

\subsection{Luminosity Segregation: Where Does it Happen?}

In Figs.~\ref{fig:gdr_radius} and~\ref{fig:gdr_density}, we demonstrate that the GDR starts to change at around $R=0.9$ Mpc or $\Sigma=40$ Mpc$^{-2}$. In Fig.~\ref{fig:radec}, the crosses denote the projected positions on the sky of cluster members in this transition region with $10<\Sigma<40$ Mpc$^{-2}$. As can be seen, this density regime corresponds to the outskirts of the cluster.  This environment coincides with that where the relative morphological population fractions and star formation rate (SFR) of galaxies starts to exhibit significant change, at least in $z<0.1$ surveys. For example, Goto et al. (2003b) studied the morphology-density relation using a large sample from the SDSS to find that galaxy morphology starts to change around the virial radius (Girardi et al. 1998), which is quite similar to the environments we found (see also Postman \& Geller 1984). Tanaka et al.(2004) studied the environmental dependence of galaxy SFR using a similar sample from the SDSS and found that the SFR starts to change at a similar environment. The passive spiral galaxies ---spiral galaxies with very little or no star formation activity --- are also found to preferentially exist in the same environment (Goto et al. 2003c; Yamauchi \& Goto 2004).  The galaxies in MS1054-0321 appear to follow similar relationship between color and density/clustocentric radius.  Figs.~\ref{fig:iz_distance} and~\ref{fig:iz_density} show the $i_{775}-z_{850}$ color of cluster galaxies as a function of $R$ and $\Sigma$. The solid line connects median values in each bin. The galaxy colors start to change significantly at around $R=$0.9 Mpc or $\Sigma=$10-40 Mpc$^{-2}$, which is the same environment where the GDR starts to change.  Postman et al. (2004) find that the morphological composition of galaxies starts to change at a similar environment at $z\sim 1$ as well.  A substantial volume of evidence thus suggests there is at least one physical mechanism that begins to become important at around $R=$0.9 Mpc or $\Sigma=$10-40 Mpc$^{-2}$ and plays a role in altering the luminosity, SFR, and morphology of cluster galaxies.

Various physical mechanisms have been proposed to explain this
cluster galaxy transformation. These mechanisms include ram-pressure stripping
 of cold gas by the intra-cluster medium (Gunn \& Gott 1972; 
 Farouki \& Shapiro 1980; Kent 1981; Fujita \& Nagashima 1999;  Abadi, Moore \& Bower 1999; 
 Quilis, Moore \& Bower 2000; Fujita \& Goto 2004);  galaxy harassment via high speed impulsive
 encounters (Moore et al. 1996, 1999; Fujita 1998); cluster
 tidal forces (Byrd \& Valtonen 1990; Valluri 1993; Fujita 1998; Gnedin 2003a,b) which
 distort galaxies as they come close to the centre; interaction/merging of
 galaxies (Icke 1985; Lavery \& Henry 1988; Mamon 1992; Makino \& Hut
  1997; Bekki 1998;  Finoguenov et al. 2003); evaporation of the cold
 gas in disc galaxies via heat conduction from the surrounding hot ICM
 (Cowie \& Songaila 1977; Fujita 2004); and a gradual decline in
 the SFR of a galaxy due to the stripping of hot halo gas (strangulation or
 suffocation; Larson, Tinsley \& Caldwell 1980; Bekki et al. 2002;
 Kodama et al. 2001b; Finoguenov et al. 2003). 

Among these processes, ram-pressure stripping and the evaporation of the 
cold gas may be disfavored by our results. Since 0.9 Mpc is almost the virial radius, the ICM density will be too 
low for these mechanisms to work with any substantial efficiency 
(Kodama et al. 2001b; Treu et al. 2003; also see Mamon et al. 2004; Moore et al. 2004). 
Galaxy-galaxy merging would be consistent with both of the above results. However, for galaxy-galaxy merging to be 
an efficient mechanism the relative velocity of the merging galaxies must be comparable with their internal velocity dispersion
(Makino \& Hut 1997). Thus, such mergers would have to take place in infalling, 
low velocity dispersion groups or at an epoch when the cluster mass
was still comparable with that on group mass scales. 
All we can conclude for now is that merging remains a viable explanation, but not necessarily the only explanation, for 
the observed trends seen in MS1054-0321 as a function of density and position.
 
 \section{Conclusion}\label{conclusion}
 
 In an effort to gain further insight into the physical processes associated with cluster galaxy formation and evolution,
 we have explored correlations between the global parameters of the galaxy luminosity function and the properties of galaxies  
 in the cluster MS1054-0321 at $z=0.83$. Our study is based on 143 spectroscopically confirmed cluster members imaged with 
 the ACS in the $V_{606},i_{775}$ and $z_{850}$ filters. Due to the wide field of view and high resolution imaging 
 capability of the ACS, we were able to study cluster LFs as a 
 function of environment, color, and morphology. Our key findings are summarized below.

\begin{figure}
\begin{center}
\includegraphics[scale=0.4]{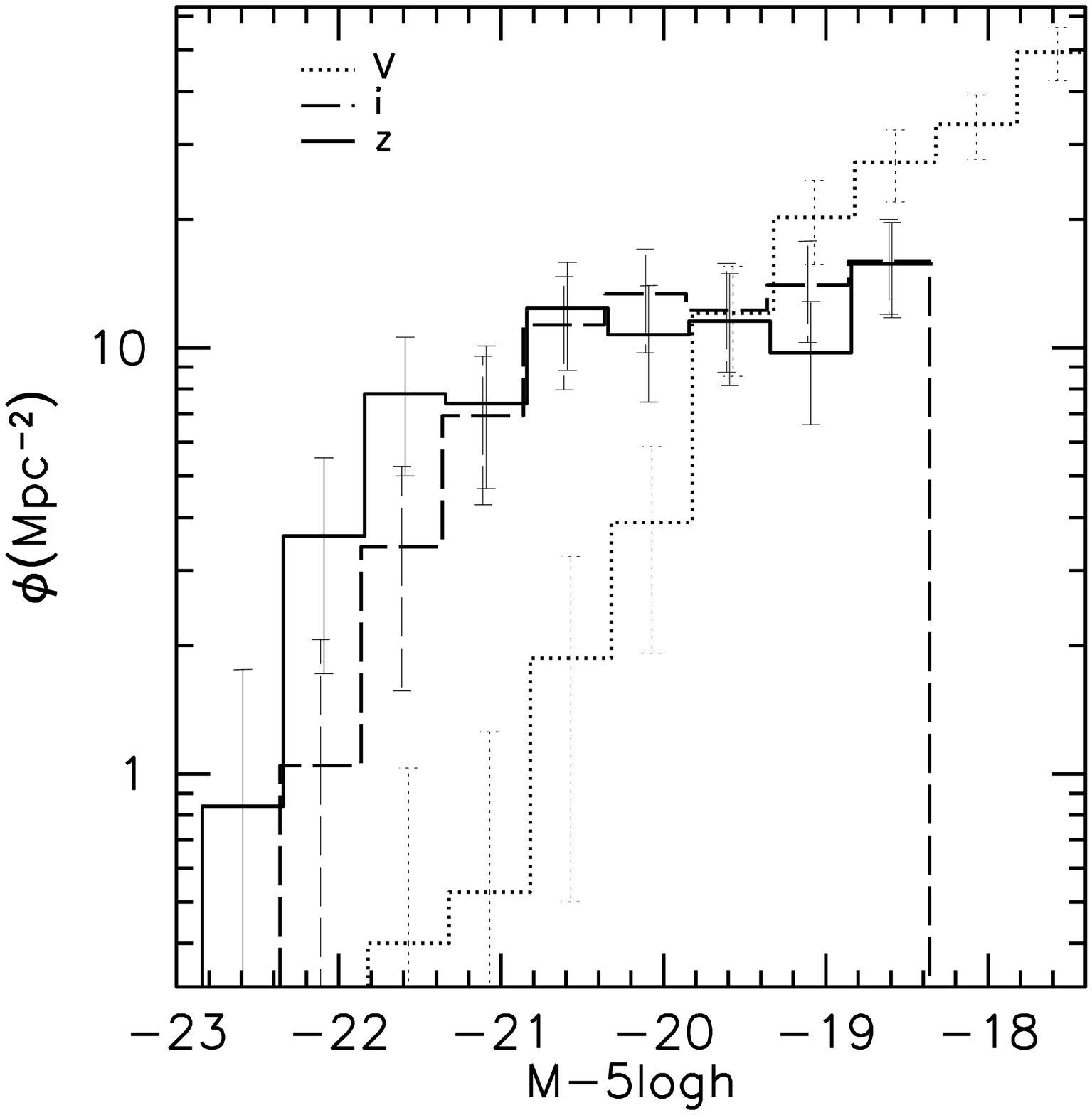}
\includegraphics[scale=0.9]{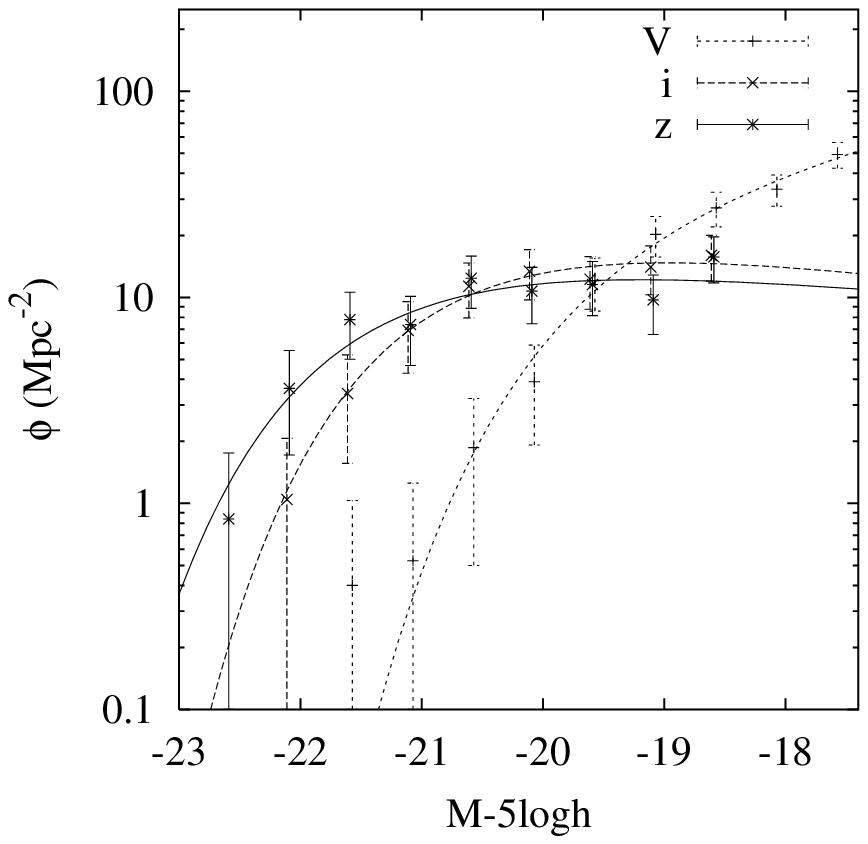}
\end{center}
\caption{The luminosity function of the spectroscopic members in the three ACS filters.
The dotted, dashed, and solid lines are for $V(F606W)$, $i(F775W)$ and $z(F850LP)$ filters in AB magnitude, respectively. The incompleteness in the spectroscopic sample is corrected using the method described in Section~\ref{correction}. The lower panel shows best-fit Schechter functions, whose parameters are presented in Table~\ref{tab:lf_parameter}. 
}\label{fig:lf_3mag}
\end{figure}

\begin{figure}
\begin{center}
\includegraphics[scale=0.4]{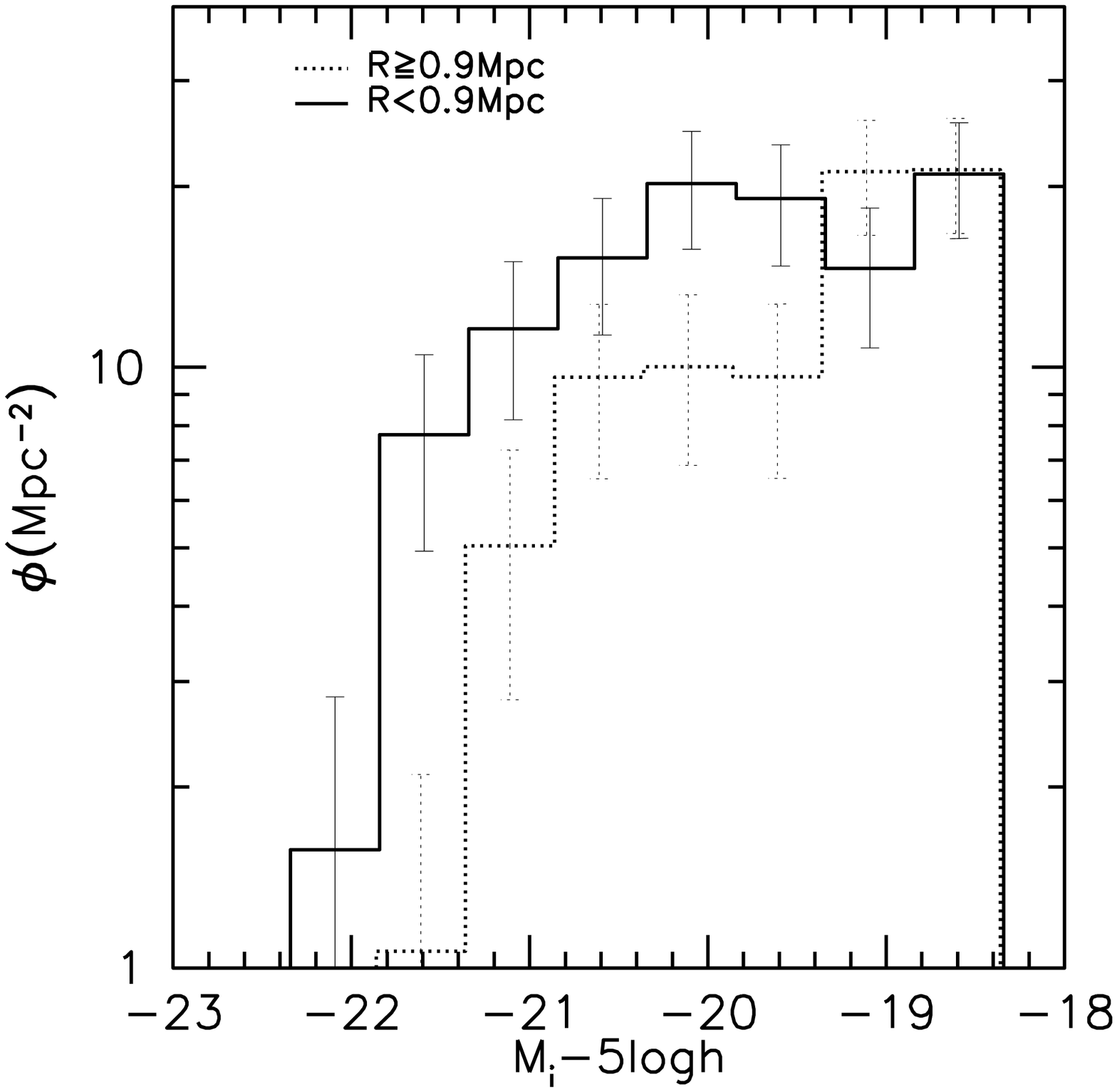}
\includegraphics[scale=0.9]{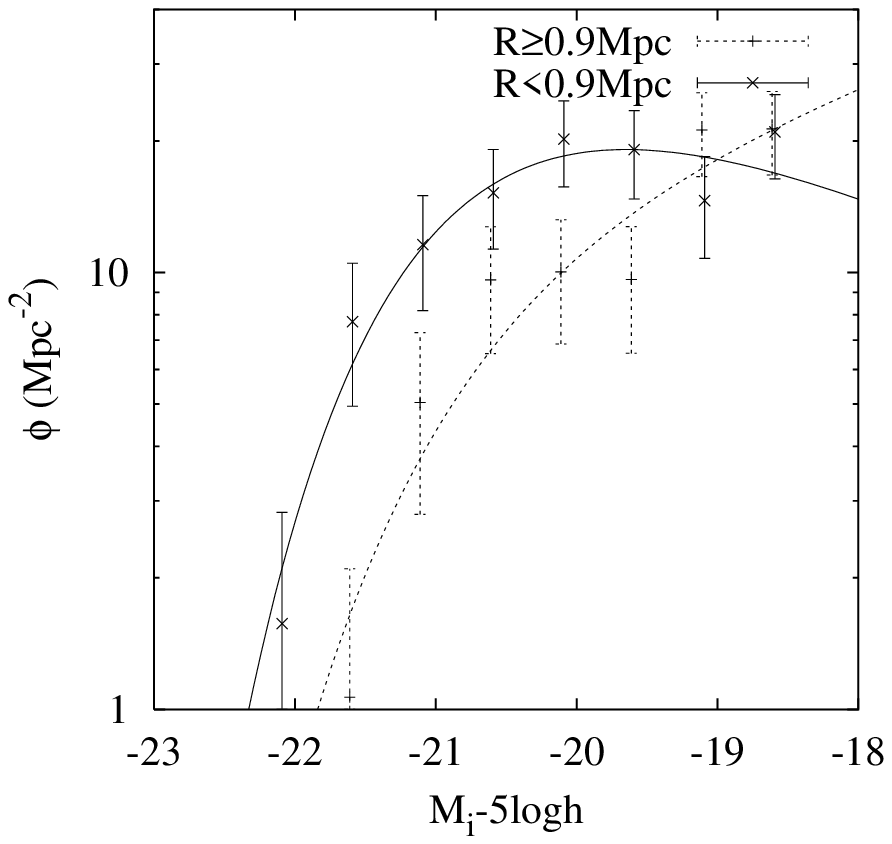}
\end{center}
\caption{The luminosity function with spectroscopic members is partitioned by the distance from the cluster center. The solid line uses galaxies within 0.9 Mpc from the cluster center determined by the position of the brightest cluster galaxy. The dotted line uses galaxies outside of 0.9 Mpc from the cluster center. 
}\label{fig:lf_distance}
\end{figure}

\begin{figure}
\begin{center}
\includegraphics[scale=0.4]{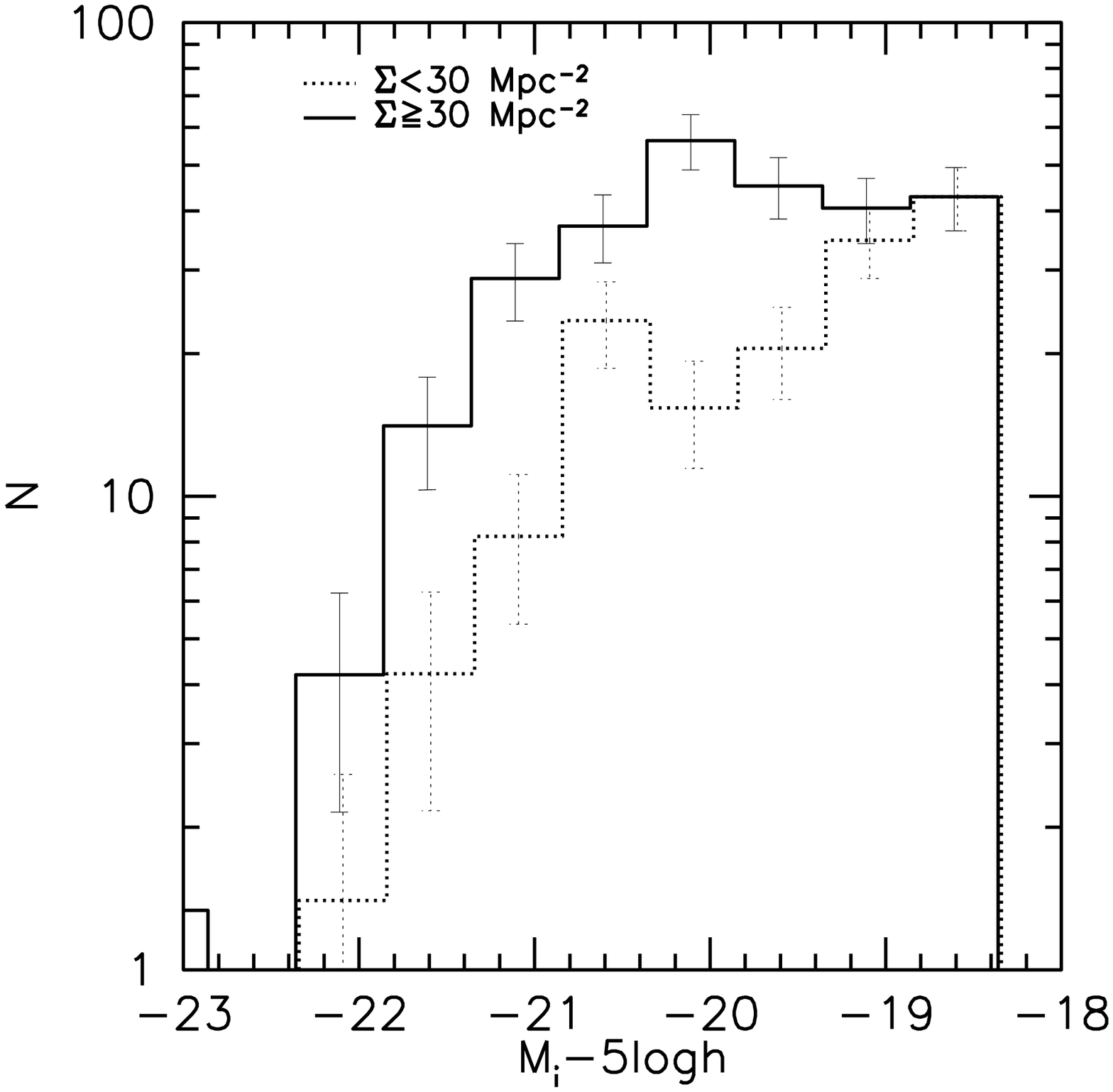}
\includegraphics[scale=0.9]{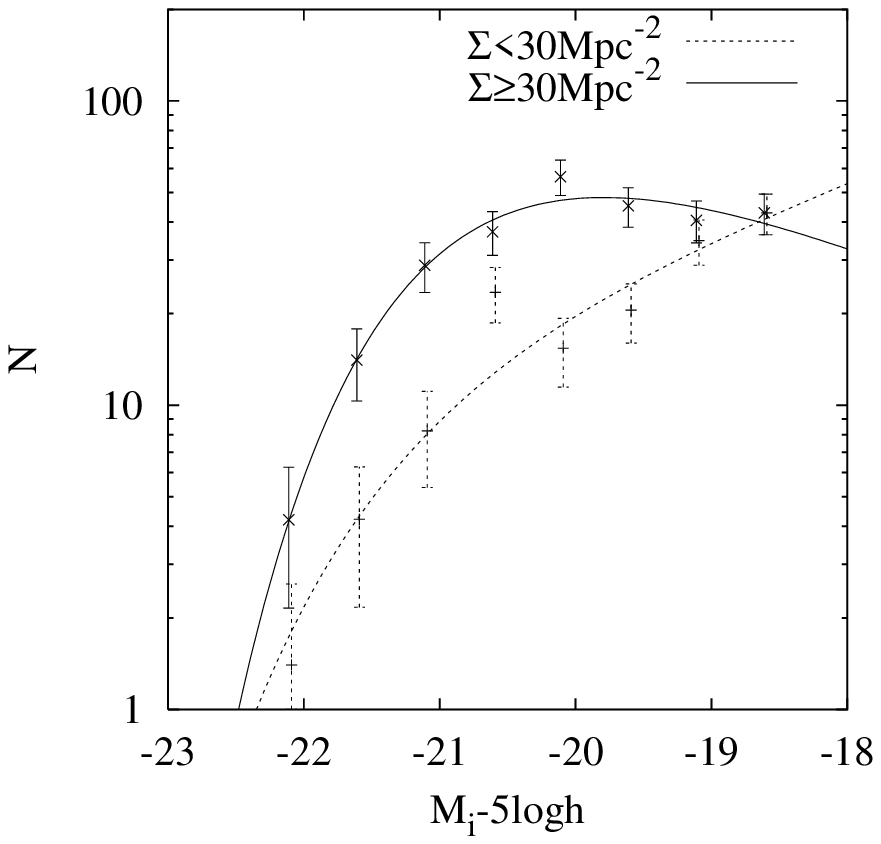}
\end{center}
\caption{ The luminosity function partitioned by the local galaxy density. The solid and dashed lines use galaxies with the local galaxy density greater than and less than 30 Mpc$^{-2}$, respectively. Since both the high density and low density samples are specially spread over the entire sky area, the same (over all) incompleteness corrections are applied to the both LFs.
}\label{fig:lf_density}
\end{figure}

\begin{figure}
\begin{center}
\includegraphics[scale=0.4]{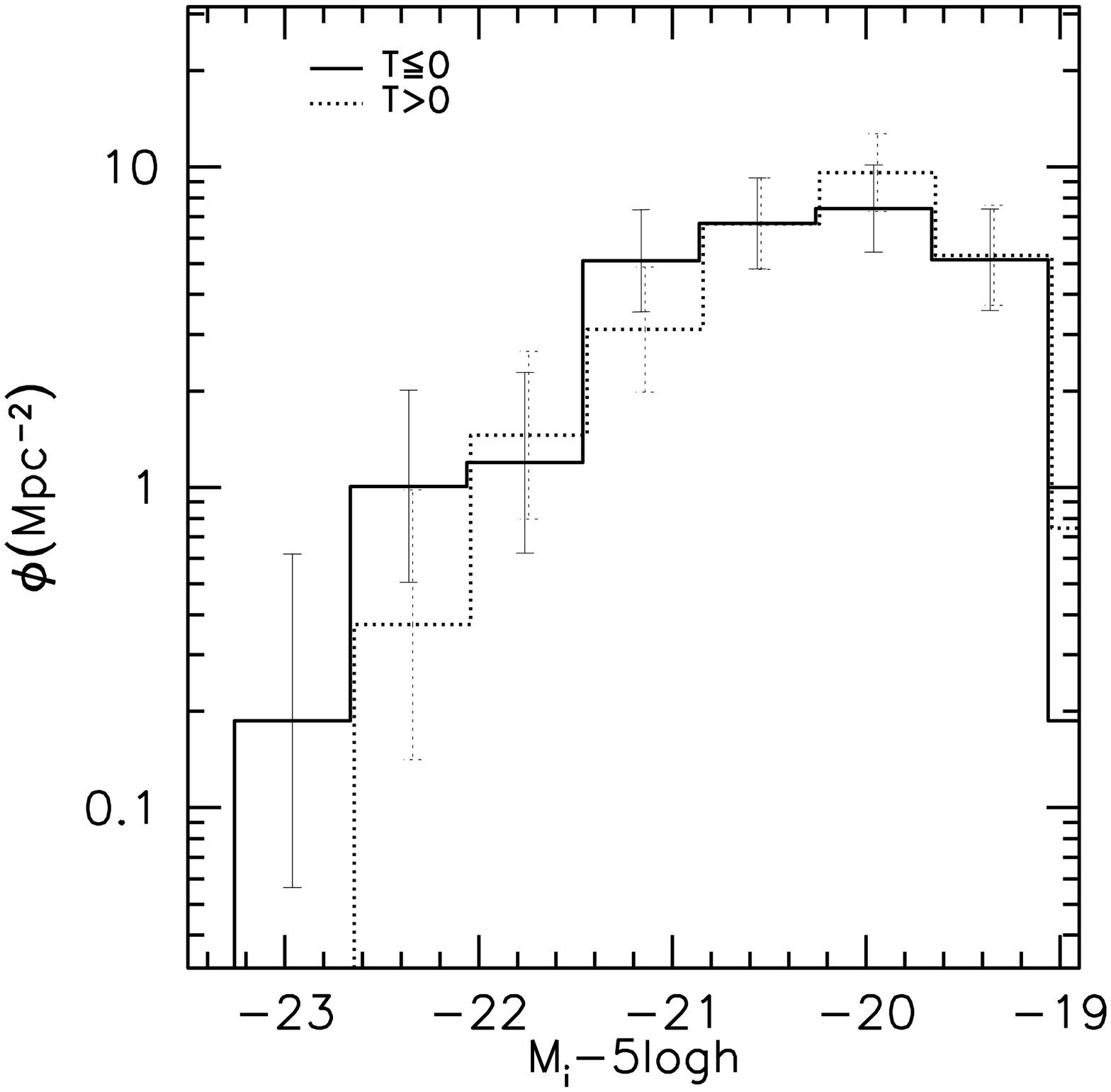}
\includegraphics[scale=0.9]{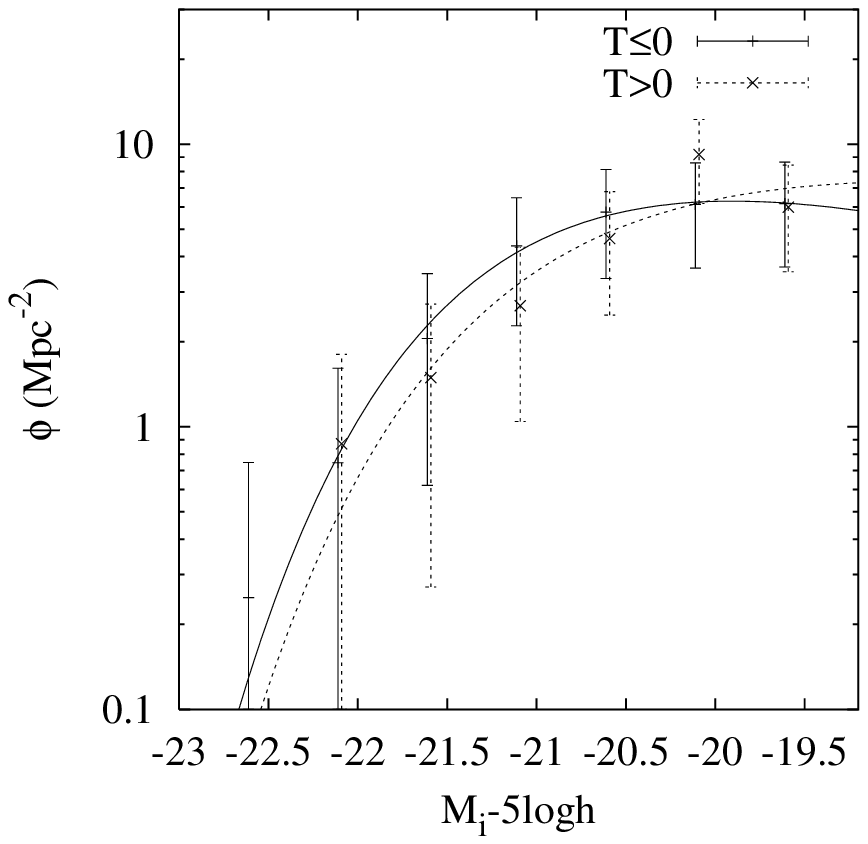}
\end{center}.
\caption{LFs divided by morphological type; $T\leq 0$:E/S0, $T>0$:late type. Corrections for incompleteness are applied separately for each morphological sample. 
}\label{fig:lf_morph}
\end{figure}

\begin{figure}
\begin{center}
\includegraphics[scale=0.8]{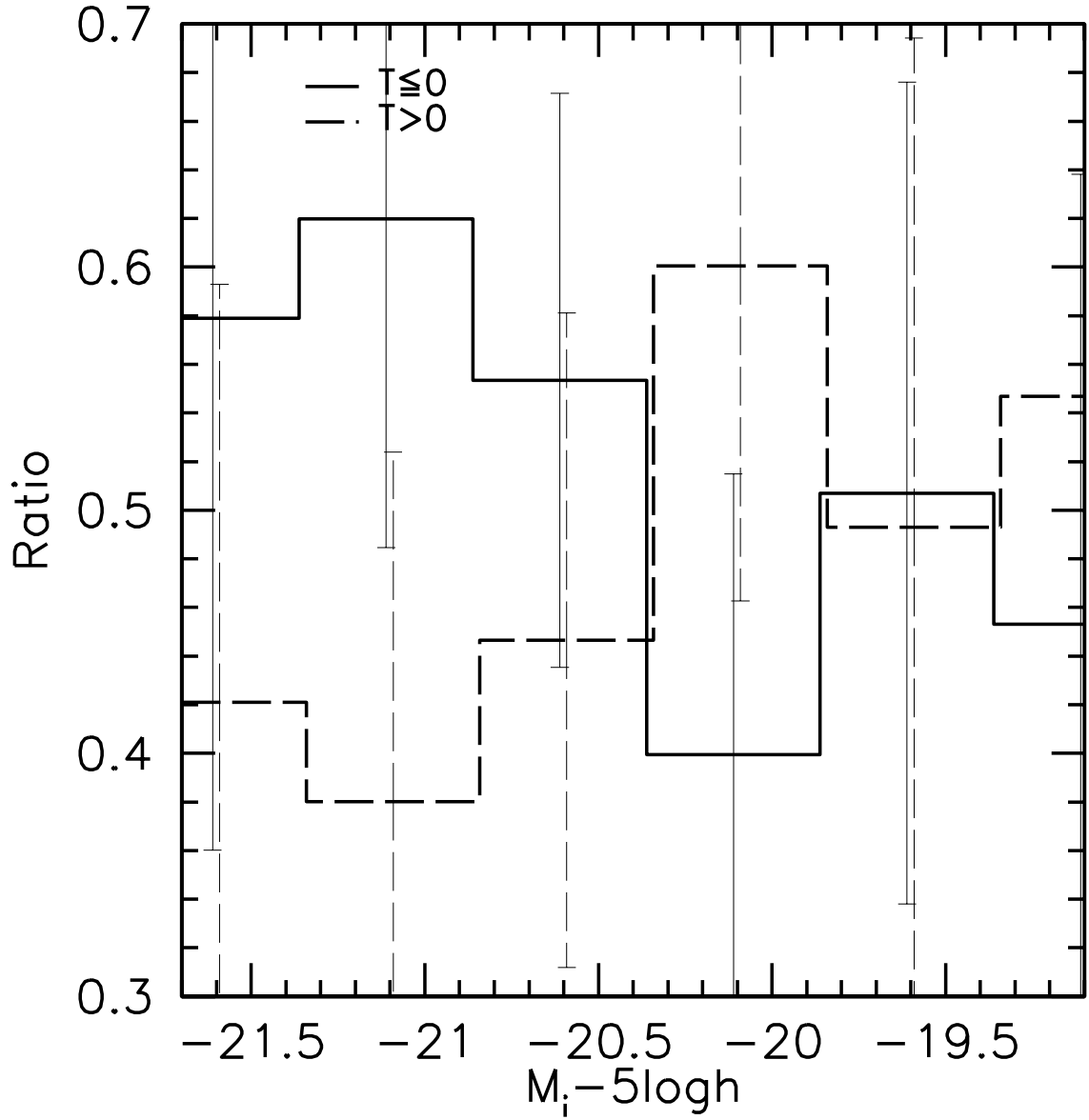}
\end{center}
\caption{
The number ratio of early-type ($T\leq 0$; solid line) and late-type ($T> 0$; dashed line) galaxies as a function of absolute magnitude. 
}\label{fig:morph_ratio}
\end{figure}

\begin{figure}
\begin{center}
\includegraphics[scale=0.8]{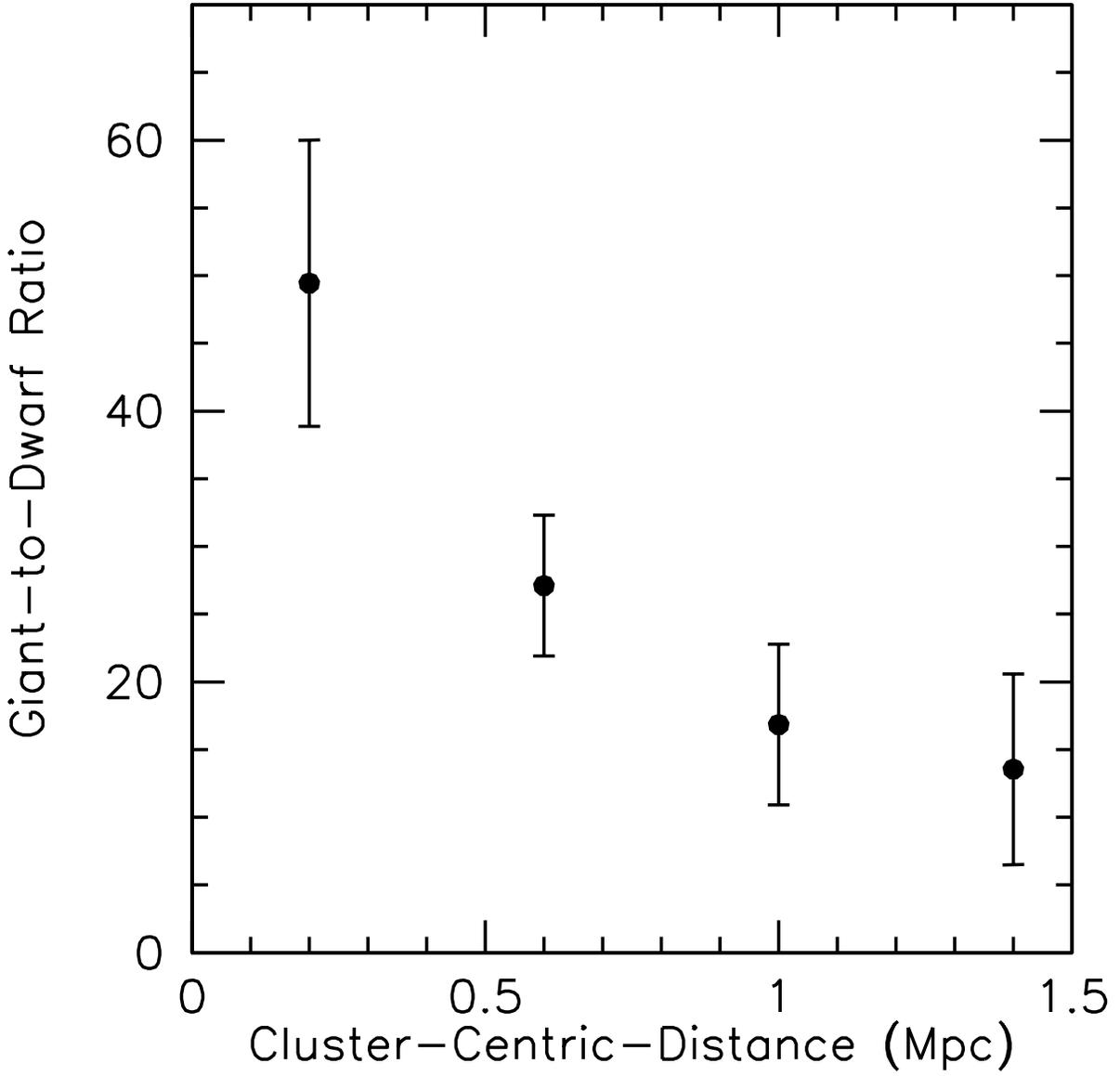}
\end{center}
\caption{Giant-to-Dwarf ratio as a function of clustocentric-radius.
}\label{fig:gdr_radius}
\end{figure}

\begin{figure}
\begin{center}
\includegraphics[scale=0.8]{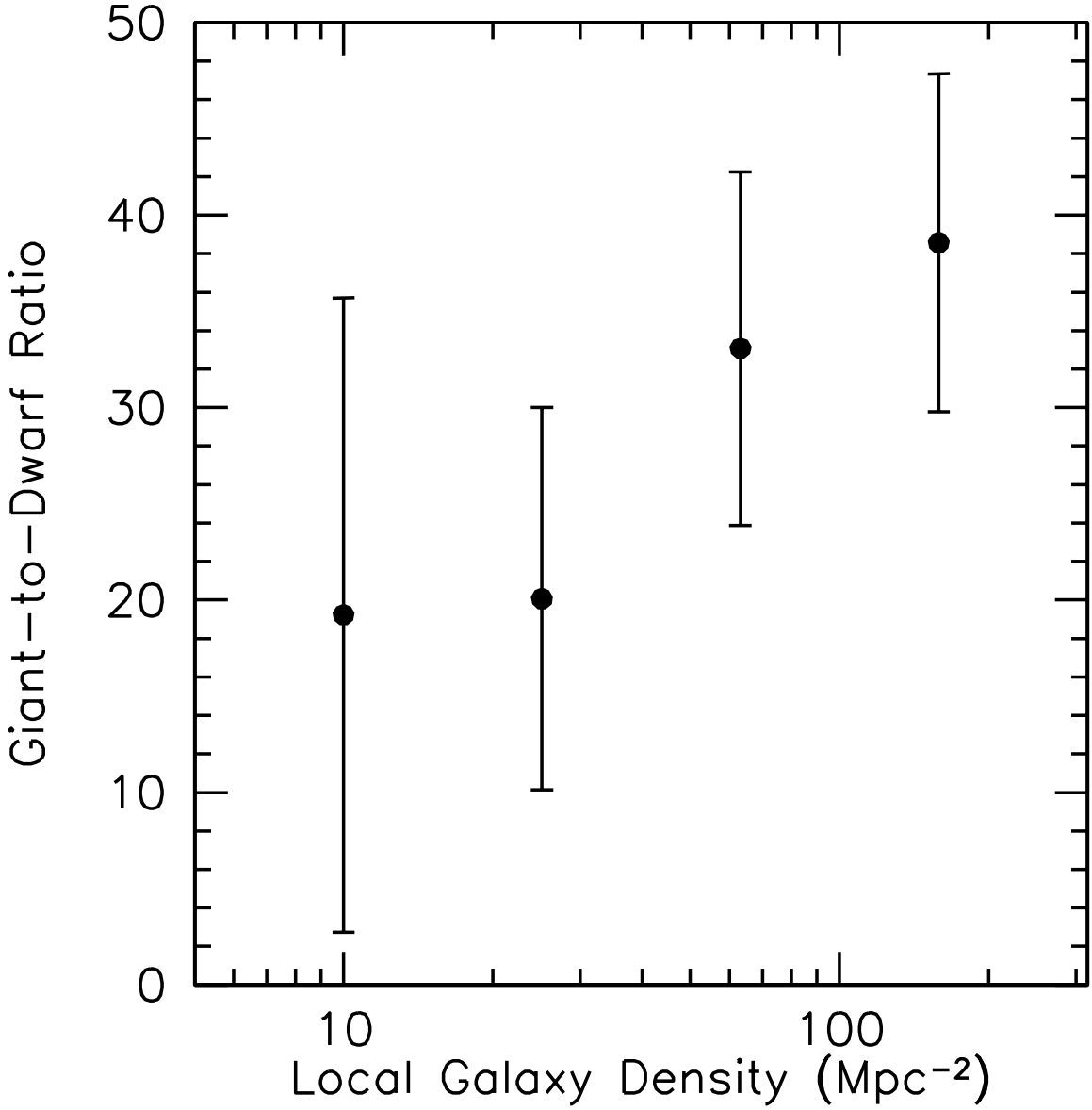}
\end{center}
\caption{Giant-to-Dwarf ratio as a function of local galaxy density.
}\label{fig:gdr_density}
\end{figure}

\begin{figure}
\begin{center}
\includegraphics[scale=0.8]{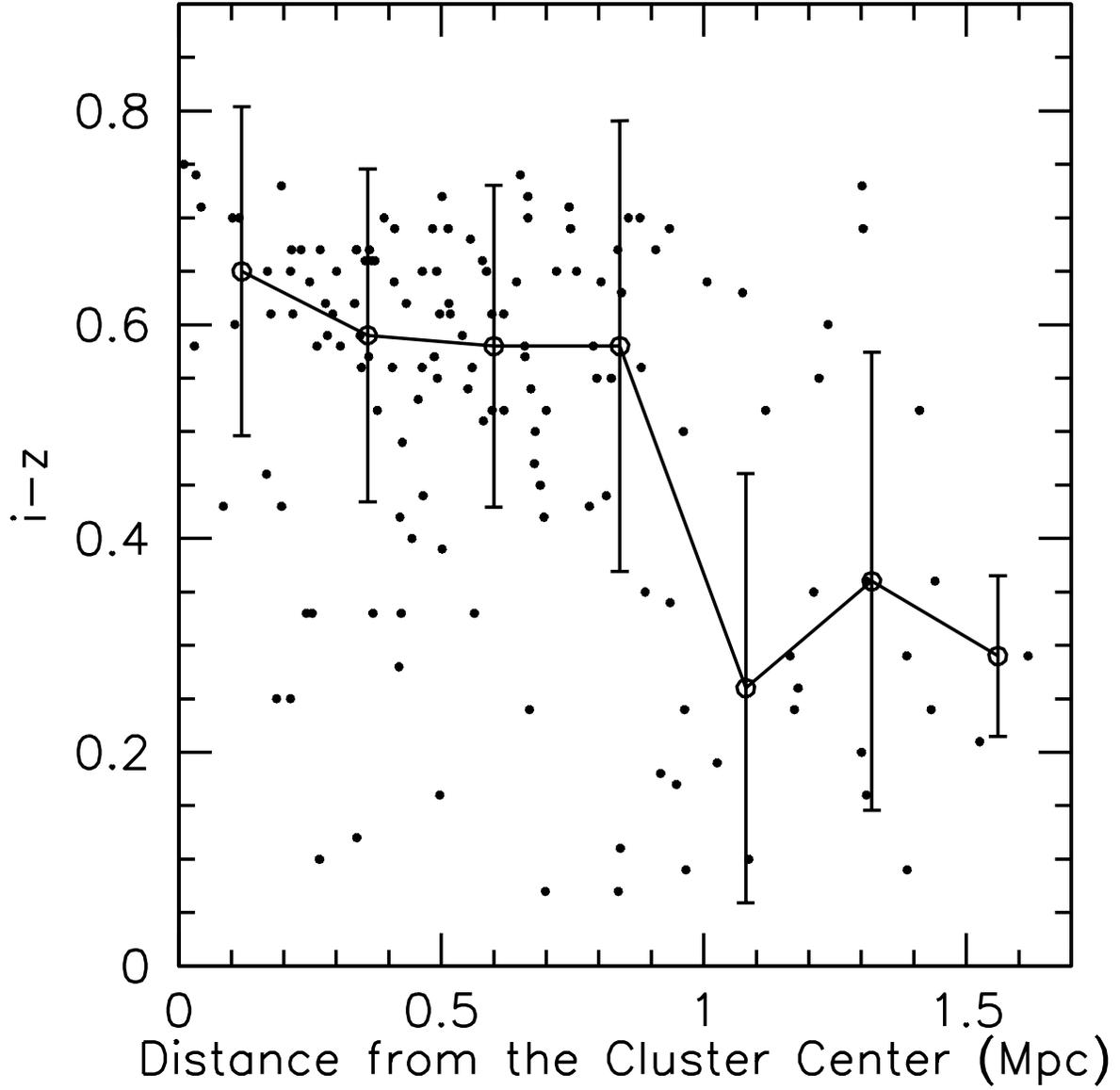}
\end{center}
\caption{The $i_{775}-z_{850}$ color is plotted against the clustocentric distance. The solid line connects medians in each bin. The error bars are based on rms in each bin.
}\label{fig:iz_distance}
\end{figure}

\begin{figure}
\includegraphics[scale=0.8]{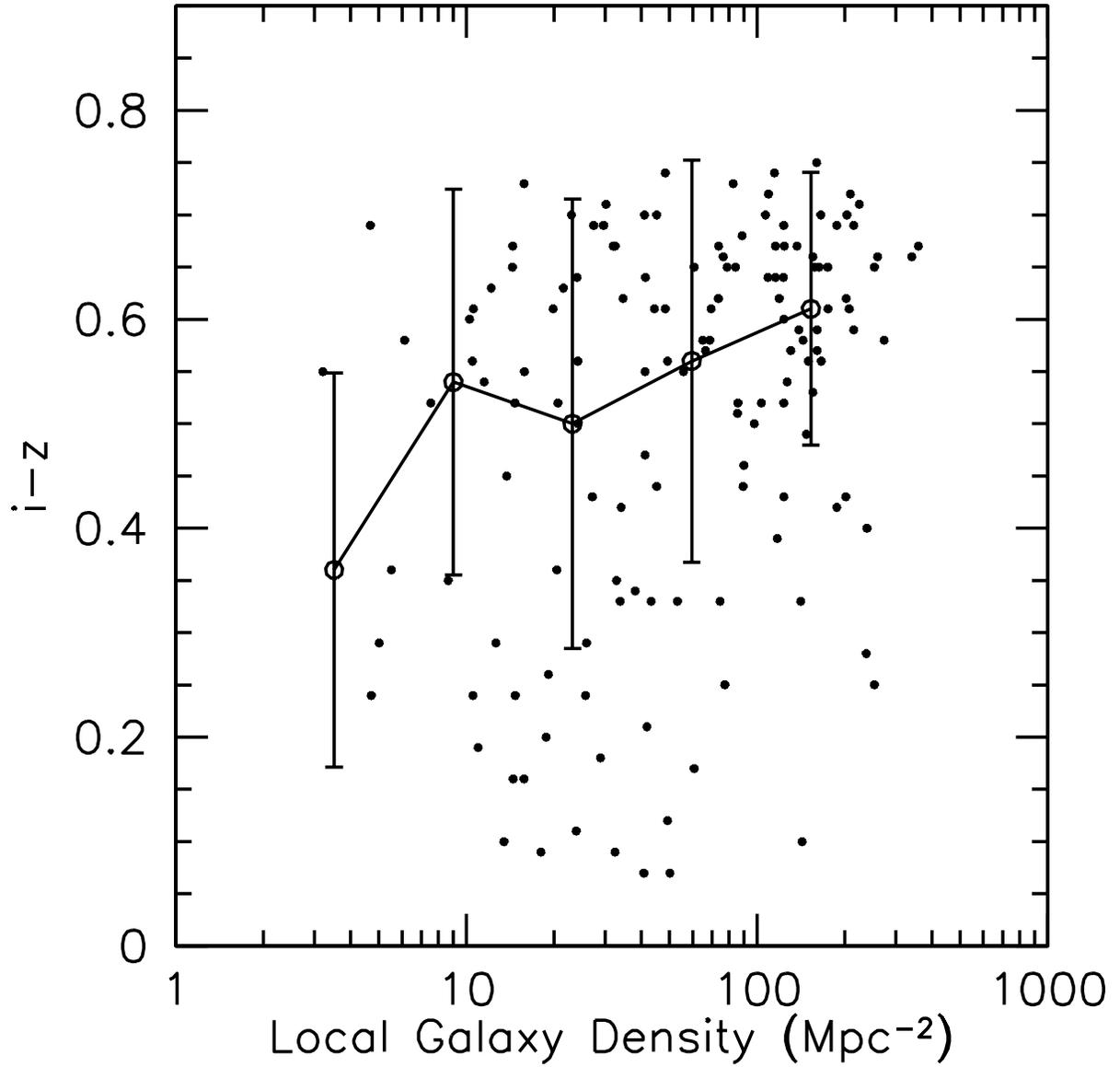}
\caption{The $i_{775}-z_{850}$ color is plotted against the local galaxy density. The solid line connects medians in each bin. The error bars are based on rms in each bin.
}\label{fig:iz_density}
\end{figure}

\begin{enumerate}
\item The faint-end slope of the LFs is steeper in the bluest band (Fig.\ref{fig:lf_3mag}).
\item The LF of galaxies in the inner part of the cluster ($R< 0.9$ Mpc) has a flatter faint-end slope 
than that in the outer regions  ($R\geq 0.9$ Mpc; Fig.\ref{fig:lf_distance}). 
The corresponding trend can be seen when the LF is measured 
as a function of the local galaxy density (Fig.~\ref{fig:lf_density}).
\item The fraction of early-type galaxies is largest at the bright end of the LF 
and gradually decreases toward fainter magnitudes (Figs.~\ref{fig:lf_morph} and~\ref{fig:morph_ratio}). 
\item The trends seen in (1), (2), and (3) are generally in agreement with observations of local galaxy clusters showing 
that the cluster core regions are dominated by bright early-type galaxies. Hence, a LF similar to what is seen at low redshift is
well established in MS1054-0321, suggesting that, on average, the galaxies in this massive cluster began their 
formation at redshifts considerably greater than unity. 
\item When the LF is partitioned by $i_{775}-z_{850}$ color, 
we find a marginally significant ($\sim2\sigma$) deficit of faint red galaxies ($i_{775}-z_{850}\geq 0.5$, $M_i>-19$). 
The deficit is inferred as well when we analyze the LF as a function of morphology -- a possible lack of early-type galaxies 
in the faintest bin (Fig.~\ref{fig:lf_morph}).  If real, these trends may suggest that faint, red, early-type galaxies 
have not been yet created in MS1054-0321. A deficit of such galaxies has not been seen in low-$z$ clusters.
\item The giant-to-dwarf ratio exhibits a strong radial dependence beginning at around the virial radius or a strong density dependence when the density increases above $\Sigma\sim$30 Mpc$^{-2}$.  The ratio is higher in the inner parts (or higher density regions) of the cluster (Figs.~\ref{fig:gdr_radius} and~\ref{fig:gdr_density}). This environment roughly corresponds to the same location where galaxy color and the morphology start to vary significantly with position and density (Figs.~\ref{fig:iz_distance} and~\ref{fig:iz_density}).  A physical process effective at this environment, such as galaxy merging, may be responsible for the cluster galaxy evolution in these transition regions.
 \end{enumerate}

\acknowledgments
We thank the anonymous referee for providing constructive and insightful comments that 
have greatly improved the clarity of this paper.
We also thank Pieter G. van Dokkum for very useful comments. The
ACS was developed under NASA contract NAS 5-32865, and this research is
supported by NASA grant NAG5-7697. Some of the results in this paper were 
based on computations performed on equipment generously provided through
a grant from Sun Microsystems, Inc. 

Facilities: \facility{HST(ACS)}, \facility{Keck(LRIS)}.

\appendix

\clearpage

\clearpage

\clearpage

\clearpage

\clearpage
\clearpage

\clearpage

\clearpage

\clearpage

\clearpage

\clearpage

\clearpage

\clearpage

\clearpage

\clearpage

\clearpage

\clearpage
\begin{table}
\caption{
 Best-fit Schechter parameters of the luminosity functions. $M^*$ is in $i_{775}$ unless otherwise specified.
}\label{tab:lf_parameter}
\begin{center}
\begin{tabular}{lrrr}
\hline
 & $M^*$ & $\alpha$ & $\chi^2$\\
\hline
All $V(F606W) $ & $-$19.59$\pm$0.24 & $-$1.35$\pm$0.13 & 0.25\\
All $i(F775W) $ & $-$20.83$\pm$0.16 & $-$0.82$\pm$0.10 & 0.12\\
All $z(F850LP)$ & $-$21.47$\pm$0.29 & $-$0.87$\pm$0.15 & 0.42 \\
\hline
$R\leq 0.9$ Mpc & $-$20.77$\pm$0.24 & $-$0.64$\pm$0.17 & 0.48\\
$R> 0.9$    Mpc & $-$20.92$\pm$0.76 & $-$1.28$\pm$0.35 & 0.97\\
\hline
$\Sigma\geq 30$ Mpc$^{-2}$ & $-$20.63$\pm$0.15 & $-$0.53$\pm$0.12   & 0.53 \\ 
$\Sigma< 30$ Mpc$^{-2}$ & $-$21.43$\pm$0.71 & $-$1.42$\pm$0.25   &  1.35\\ 
\hline
$i_{775}-z_{850} \geq 0.5$   & $-$20.11$\pm$0.22 & $-$0.09$\pm$0.27 &  0.34  \\
$i_{775}-z_{850} <0.5$       & $-$21.92$\pm$0.86 & $-$1.48$\pm$0.23 &  0.18  \\
\hline
$T \leq 0$   & $-$20.76$\pm$0.12 & $-$0.54$\pm$0.13  & 0.02  \\ 
$T > 0$       & $-$20.79$\pm$0.79 & $-$0.84$\pm$0.67  & 0.47  \\ 
\hline
\end{tabular}
\end{center}
\end{table}

\end{document}